\documentclass[final,twoside,11pt]{entics} 

\usepackage{graphicx}
\usepackage[all]{xy}
\newcommand{\hide}[1]{}

\usepackage{amsmath,mathtools,amsfonts,latexsym,stmaryrd,mathrsfs} 
\usepackage{cite}

\usepackage{enumitem}
\usepackage{longtable}

\usepackage{orcidlink} 

\SetSymbolFont{stmry}{bold}{U}{stmry}{m}{n} 
\usepackage{ebproof}
\usepackage{forest}
\usepackage{nicematrix}
\usepackage{xfrac}

\usepackage{nameref}
\usepackage{float}
\usepackage{xifthen}
\usepackage{xcolor} 
\usepackage[nameinlink]{cleveref}
\usepackage{scalerel} 

\usepackage{todonotes}
\usepackage{soul} 
\usepackage{xargs} 
\usepackage{ifmtarg} 

\usepackage{subfiles}
\usepackage{fancyhdr}
\usepackage{upgreek}

\usepackage{cases}
\usepackage{parcolumns}
\usepackage{listings}
\usepackage{multicol}
\usepackage{subfig}
\usepackage{multirow}
\usepackage{braket} 

\usepackage{quiver}
\usepackage{adjustbox} 

\usepackage{listings}

\definecolor{mygreen}{rgb}{0,0.4,0}
\definecolor{myblue}{rgb}{0,0,0.7}
\definecolor{myred}{rgb}{0.7,0,0}
\definecolor{mygray}{rgb}{0.5,0.5,0.5}
\definecolor{mymauve}{rgb}{0.58,0,0.82}

\newcommand*{\mlstinline}[1]{\mbox{\lstinline[mathescape]|#1|}}
\usepackage{newunicodechar}

\DeclareFontFamily{U}{rcjhbltx}{}
\DeclareFontShape{U}{rcjhbltx}{m}{n}{<->rcjhbltx}{}
\DeclareSymbolFont{hebrewletters}{U}{rcjhbltx}{m}{n}

\let\aleph\relax\let\beth\relax
\let\gimel\relax\let\daleth\relax

\DeclareMathSymbol{\aleph}{\mathord}{hebrewletters}{39}
\DeclareMathSymbol{\beth}{\mathord}{hebrewletters}{98}
\DeclareMathSymbol{\gimel}{\mathord}{hebrewletters}{103}
\DeclareMathSymbol{\daleth}{\mathord}{hebrewletters}{100}

\DeclareMathSymbol{\lamed}{\mathord}{hebrewletters}{108}
\DeclareMathSymbol{\mem}{\mathord}{hebrewletters}{109}
\DeclareMathSymbol{\ayin}{\mathord}{hebrewletters}{96}
\DeclareMathSymbol{\tsadi}{\mathord}{hebrewletters}{118}
\DeclareMathSymbol{\qof}{\mathord}{hebrewletters}{113}
\DeclareMathSymbol{\shin}{\mathord}{hebrewletters}{152}

\usepackage[T4,T1]{fontenc}
\newunicodechar{מ}{\mem}
\newunicodechar{ƒ}{\text{\fontencoding{T4}\selectfont\m f}}
\newunicodechar{ŋ}{\text{\fontencoding{T1}\ng{}}}

\DeclareFontFamily{U}{min}{}
\DeclareFontShape{U}{min}{m}{n}{<-> udmj30}{}
\newunicodechar{ろ}{\mathop{\text{\usefont{U}{min}{m}{n}\symbol{'215}}}}
\newunicodechar{よ}{\mathop{\text{\usefont{U}{min}{m}{n}\symbol{'210}}}}

\newunicodechar{：}{\colon}
\newunicodechar{∀}{\forall}
\newunicodechar{∃}{\exists}
\newunicodechar{α}{\alpha}
\newunicodechar{β}{\beta}
\newunicodechar{γ}{\gamma}
\newunicodechar{δ}{\delta}
\newunicodechar{ε}{\epsilon}
\newunicodechar{ζ}{\zeta}
\newunicodechar{η}{\eta}
\newunicodechar{θ}{\theta}
\newunicodechar{ϑ}{\vartheta}
\newunicodechar{ι}{\iota}
\newunicodechar{κ}{\kappa}
\newunicodechar{λ}{\lambda}
\newunicodechar{μ}{\mu}
\newunicodechar{ν}{\nu}
\newunicodechar{ξ}{\xi}
\newunicodechar{ρ}{\rho}
\newunicodechar{σ}{\sigma}
\newunicodechar{τ}{\tau}
\newunicodechar{υ}{\upsilon}
\newunicodechar{φ}{\phi}
\newunicodechar{χ}{\chi}
\newunicodechar{ψ}{\psi}
\newunicodechar{ω}{\omega}
\newunicodechar{Σ}{\Sigma}
\newunicodechar{π}{\pi}
\newunicodechar{Π}{\Pi}
\newunicodechar{Γ}{\Gamma}
\newunicodechar{Δ}{\Delta}
\newunicodechar{Θ}{\Theta}
\newunicodechar{Λ}{\Lambda}
\newunicodechar{Ξ}{\Xi}
\newunicodechar{Ψ}{\Psi}
\newunicodechar{Ω}{\Omega}
\newunicodechar{∂}{\partial}
\newunicodechar{∇}{\nabla}
\newunicodechar{Φ}{\Phi}
\newunicodechar{⊢}{\vdash}
\newunicodechar{∈}{\in}
\newunicodechar{∉}{\notin}
\newunicodechar{∋}{\ni}
\newunicodechar{∌}{\not\ni}
\newunicodechar{∅}{\emptyset}
\newunicodechar{∪}{\cup}
\newunicodechar{∩}{\cap}
\newunicodechar{∧}{\wedge}
\newunicodechar{∨}{\vee}
\newunicodechar{⊥}{\bot}
\newunicodechar{⊤}{\top}
\newunicodechar{≤}{\leq}
\newunicodechar{≥}{\geq}
\newunicodechar{≠}{\neq}
\newunicodechar{≡}{\equiv}
\newunicodechar{≈}{\approx}
\newunicodechar{≃}{\simeq}
\newunicodechar{≅}{\cong}
\newunicodechar{⊆}{\subseteq}
\newunicodechar{⊇}{\supseteq}
\newunicodechar{⊂}{\subset}
\newunicodechar{⊃}{\supset}
\newunicodechar{⟶}{\rightarrow}
\newunicodechar{→}{\to}
\newunicodechar{↦}{\mapsto}
\newunicodechar{⟼}{\mapsto}
\newunicodechar{↔}{\leftrightarrow}
\newunicodechar{↪}{\hookrightarrow}
\newunicodechar{⟵}{\leftarrow}
\newunicodechar{↦}{\mapsto}
\newunicodechar{⟹}{\Longrightarrow}
\newunicodechar{⇒}{\Rightarrow}
\newunicodechar{⟸}{\Longleftarrow}
\newunicodechar{⟺}{\Leftrightarrow}
\newunicodechar{⊗}{\otimes}
\newunicodechar{×}{\times}
\newunicodechar{∞}{\infty}
\newunicodechar{…}{\ldots}
\newunicodechar{⋯}{\cdots}
\newunicodechar{⋮}{\vdots}
\newunicodechar{⋱}{\ddots}
\newunicodechar{⋰}{\ddots}
\newunicodechar{⋅}{\cdot}
\newunicodechar{∧}{\wedge}
\newunicodechar{∨}{\vee}
\newunicodechar{∩}{\cap}
\newunicodechar{∪}{\cup}
\newunicodechar{⨆}{\sqcup}
\newunicodechar{¬}{\neg}
\newunicodechar{ℙ}{\mathbb{P}}
\newunicodechar{ℕ}{\mathbb{N}}
\newunicodechar{ℝ}{\mathbb{R}}
\newunicodechar{𝕍}{\mathbb{V}}
\newunicodechar{𝕊}{\mathbb{S}}
\newunicodechar{𝕄}{\mathbb{M}}
\newunicodechar{𝔸}{\mathbb{A}}
\newunicodechar{𝔹}{\mathbb{B}}
\newunicodechar{𝔻}{\mathbb{D}}
\newunicodechar{𝔼}{\mathbb{E}}
\newunicodechar{𝔽}{\mathbb{F}}
\newunicodechar{≝}{\defeq}
\newunicodechar{𝒞}{\mathscr{C}}
\newunicodechar{𝒟}{\mathscr{D}}
\newunicodechar{𝒢}{\mathscr{G}}
\newunicodechar{𝒥}{\mathscr{J}}
\newunicodechar{𝒦}{\mathscr{K}}
\newunicodechar{𝒩}{\mathscr{N}}
\newunicodechar{𝒪}{\mathscr{O}}
\newunicodechar{𝒫}{\mathscr{P}}
\newunicodechar{𝒬}{\mathscr{Q}}
\newunicodechar{𝒮}{\mathscr{S}}
\newunicodechar{𝒯}{\mathscr{T}}
\newunicodechar{𝒰}{\mathscr{U}}
\newunicodechar{𝒱}{\mathscr{V}}
\newunicodechar{𝒲}{\mathscr{W}}
\newunicodechar{𝒳}{\mathscr{X}}
\newunicodechar{𝒴}{\mathscr{Y}}
\newunicodechar{𝒵}{\mathscr{Z}}
\newunicodechar{𝖌}{\mathfrak{g}}
\newunicodechar{𝖍}{\mathfrak{h}}
\newunicodechar{𝖎}{\mathfrak{i}}
\newunicodechar{𝖏}{\mathfrak{j}}
\newunicodechar{𝖐}{\mathfrak{k}}
\newunicodechar{𝖑}{\mathfrak{l}}
\newunicodechar{𝖒}{\mathfrak{m}}
\newunicodechar{𝖓}{\mathfrak{n}}
\newunicodechar{𝖔}{\mathfrak{o}}
\newunicodechar{𝖕}{\mathfrak{p}}
\newunicodechar{𝖖}{\mathfrak{q}}
\newunicodechar{𝖗}{\mathfrak{r}}
\newunicodechar{𝖘}{\mathfrak{s}}
\newunicodechar{𝖙}{\mathfrak{t}}
\newunicodechar{𝖚}{\mathfrak{u}}
\newunicodechar{𝖛}{\mathfrak{v}}
\newunicodechar{𝖜}{\mathfrak{w}}
\newunicodechar{𝖝}{\mathfrak{x}}
\newunicodechar{𝖞}{\mathfrak{y}}
\newunicodechar{𝖟}{\mathfrak{z}}
\newunicodechar{𝕬}{\mathcal{A}}
\newunicodechar{𝕭}{\mathcal{B}}
\newunicodechar{𝕮}{\mathcal{C}}
\newunicodechar{𝕯}{\mathcal{D}}
\newunicodechar{𝕰}{\mathcal{E}}
\newunicodechar{𝕱}{\mathcal{F}}
\newunicodechar{𝕲}{\mathcal{G}}
\newunicodechar{𝕳}{\mathcal{H}}
\newunicodechar{𝕴}{\mathcal{I}}
\newunicodechar{𝕵}{\mathcal{J}}
\newunicodechar{𝕶}{\mathcal{K}}
\newunicodechar{𝕷}{\mathcal{L}}
\newunicodechar{𝕸}{\mathcal{M}}
\newunicodechar{𝕹}{\mathcal{N}}
\newunicodechar{𝕺}{\mathcal{O}}
\newunicodechar{𝕻}{\mathcal{P}}
\newunicodechar{𝕼}{\mathcal{Q}}
\newunicodechar{𝕽}{\mathcal{R}}
\newunicodechar{𝕾}{\mathcal{S}}
\newunicodechar{𝕿}{\mathcal{T}}
\newunicodechar{𝖀}{\mathcal{U}}
\newunicodechar{𝖁}{\mathcal{V}}
\newunicodechar{𝖂}{\mathcal{W}}
\newunicodechar{𝖃}{\mathcal{X}}
\newunicodechar{𝖄}{\mathcal{Y}}
\newunicodechar{𝖅}{\mathcal{Z}}

\newcommand{\cb}{\color{myblue}}

\newcommand{\id}{\mathrm {id}}

\newcommand{\x}{\mathbf{x}}

\newcommand{\BiGrphemb}{\mathbf{BiGrph}_{\scalebox{0.9}[1]{\scriptsize \textit{emb}}}}

\newcommand{\Var}{\mathsf{Var}}

\newcommand{\op}{^{\mathrm{op}}}

\newcommand{\Atoms}{\mathbb{A}}

\newcommand{\defeq}{\stackrel{\mbox{\tiny{def}}}=}

\newcommand{\Sets}{\mathbf{Set}} 

\newcommand{\NN}{\mathbb N}

\newcommand{\tbool}{\mathsf{bool}}
\newcommand{\ifalse}{\mathsf{false}}
\newcommand{\itrue}{\mathsf{true}}
\newcommand{\ite}[3]{\mathsf{if}\,#1\,\mathsf{then}\,#2\,\mathsf{else}\,#3}
\newcommand{\imatch}[3]{\mathsf{match}\,#1\,\mathsf{as}\,#2\,\mathsf{in}\,#3}
\newcommand{\ireturn}{\mathsf{return}}
\newcommand{\iletin}[3]{\mathsf{let~val}~#1~\leftarrow~#2\mathsf{\,in\,}#3}
\newcommand{\ifresh}{\mathsf{fresh}()}
\newcommand{\iflip}[1][θ]{\mathsf{flip}(#1)}

\makeatletter
\newcommand{\lambdamem}[2]{%
  \lambda_{\mathsf{\mem}}\@ifmtarg{#1}{}{\!\;#1.\:#2}}
\newcommand{\denpar}[2][]{\@ifmtarg{#1}{\llparenthesis #2 \rrparenthesis}{\llparenthesis #2 \rrparenthesis_{#1}}}
\newcommand{\den}[2][]{\@ifmtarg{#1}{\llbracket #2 \rrbracket}{\llbracket #2 \rrbracket_{#1}}}
\makeatother

\newcommand{\Closures}{\mathsf{Closures}}

\newcommand{\Tree}[1][2 + 𝖌_L + 𝖌_R]{\mathop{\mathsf{Tree}}(#1)}

\newcommandx*{\memobraces}[3][2={ƒ, a}, 3=γ]{\left\{\!\!\left\{#1\right\}\!\!\right\}^{#2}_{#3}}
\newcommandx*{\bigmemobraces}[3][2={ƒ, a}, 3=γ]{\big\{\!\!\big\{#1\big\}\!\!\big\}^{#2}_{#3}}
\newcommandx*{\Bigmemobraces}[3][2={ƒ, a}, 3=γ]{\Big\{\!\!\Big\{#1\Big\}\!\!\Big\}^{#2}_{#3}}

\renewcommand{\tt}{\mathbf{tt}}

\newcommand{\C}{{\mathcal C}}
\newcommand{\J}{\mathop{\mathrm{J}}}

\newcommand{\letin}[3]{\mathrm{let}~#1~=~#2\mathrm{\,in\,}#3}
\newcommand{\for}[1]{\mathrm{for~each}~#1,\;}
\newcommand{\lambdaabs}[2]{\lambda\!\;#1.\:#2}
\newcommand{\fv}{\mathsf{fv}}

\newcommand{\rvline}{\hspace*{-\arraycolsep}\vline\hspace*{-\arraycolsep}}

\newcommand{\vj}[3]{#1\mathrel{\vdash\!\!\!\!^{\mathsf{v}}} #2:#3}
\newcommand{\cj}[3]{#1\mathrel{\vdash\!\!\!\!^{\mathsf{c}}} #2:#3}
\newcommand{\cje}[4]{#1 \mid #2\mathrel{\vdash\!\!\!\!^{\mathsf{c}}} #3:#4}

\newcommand{\Fns}{\mathbb F}

\newcommand{\BiG}{\BiGrphemb}
\newcommand{\colim}{\mathop {\mathrm{colim}}}
\newcommand{\Prfin}{P\!_{\mathrm{f}}}

\newcommand{\bind}{\mathop{\gg\mkern-10mu\scalebox{1}[1]{=}}}

\definecolor{cambridgeblue}{RGB}{163, 193, 173}
\definecolor{babyblueeyes}{rgb}{0.63, 0.79, 0.95}
\definecolor{purple}{RGB}{128,0,128}
\definecolor{orchid}{RGB}{218,112,214}
\definecolor{crimson}{RGB}{220,20,60}
\definecolor{salmon}{RGB}{250,128,114}
\definecolor{brown}{RGB}{165,42,42}
\definecolor{chocolate}{RGB}{210,105,30}
\definecolor{sandybrown}{RGB}{244,164,96}
\definecolor{darkorange}{RGB}{255,140,0}
\definecolor{khaki}{RGB}{240,230,140}
\definecolor{greenyellow}{RGB}{173,255,47}
\definecolor{lightseagreen}{RGB}{32,178,170}
\definecolor{mediumseagreen}{RGB}{60,179,113}
\definecolor{olivedrab}{RGB}{107,142,35}
\definecolor{teal}{RGB}{0,128,128}
\definecolor{cadetblue}{RGB}{95,198,160}
\definecolor{aquamarine}{RGB}{127,255,212}
\definecolor{steelblue}{RGB}{70,130,180}
\definecolor{navy}{RGB}{0,0,128}
\definecolor{midnightblue}{RGB}{25,25,112}

\DeclareMathAlphabet{\mathbf}{OT1}{cmr}{b}{n}
\DeclareMathAlphabet{\mathpzc}{OT1}{pzc}{m}{it}

\newcommand{\Inj}{\mathbf{Inj}}

\newcommand{\pushout}{\coprod}

\newcommand{\xinjto}{\xhookrightarrow}
\newcommand{\xto}{\xrightarrow}

\newcommand{\then}{\mathbin ;}

\renewcommand{\vec}{\overrightarrow}

\newcommand{\eg}{\textit{e.g.}~}
\newcommand{\ie}{\textit{i.e.}~}

\newcommand{\mmid}{\; $\mid$ \;}

\usepackage{makecell}

\usepackage{enticsmacro}

\def\conf{MFPS 2023} 	
\def\myconf{mfps39}
\volume{3}			
\def\volu{3}			
\def\papno{10}				
\def\lastname{Kaddar and Staton} 
\def\runtitle{Stochastic memoization and name generation} 
\def\mydoi{12291}
\def\copynames{Y. Kaddar, S. Staton}    
\begin{document}
\begin{frontmatter}
  \title{A Model of Stochastic Memoization and Name Generation in Probabilistic Programming: \\Categorical Semantics via Monads on Presheaf Categories
  }
  \author{Younesse Kaddar}
  and \author{Sam Staton}
  \address{Department of Computer Science, University of Oxford, UK}
\begin{abstract} 
    Stochastic memoization is a higher-order construct of probabilistic programming languages that is key in Bayesian nonparametrics, a modular approach that allows us to extend models beyond their parametric limitations and compose them in an elegant and principled manner. Stochastic memoization is simple and useful in practice, but semantically elusive, particularly regarding dataflow transformations. As the naive implementation resorts to the state monad, which is not commutative, it is not clear if stochastic memoization preserves the dataflow property -- \ie whether we can reorder the lines of a program without changing its semantics, provided the dataflow graph is preserved.
    In this paper, we give an operational and categorical semantics to stochastic memoization and name generation in the context of a minimal probabilistic programming language, for a restricted class of functions.
    Our contribution is a first model of stochastic memoization of constant Bernoulli functions with a non-enumerable type, which validates data flow transformations, bridging the gap between traditional probability theory and higher-order probability models. Our model uses a presheaf category and novel probability monad on it.
\end{abstract}
\end{frontmatter}

\section{Introduction}
\label{sec:intro}
Bayesian nonparametric models are a powerful approach to statistical learning. Unlike parametric models, which have a fixed number of parameters, nonparametric models can have an unbounded number of parameters that grows as needed to fit complex data. This flexibility allows them to capture subtle patterns in data that parametric models may miss, and it makes them more composable, because they are not arbitrarily truncated.

Prominent examples of nonparametric models include Dirichlet process models for clustering similar data points, and the Infinite Relational Model for automatically discovering latent groups and features, amongst others. These infinite-dimensional models can accommodate an unbounded number of components, clusters, or other features in order to fit observed data as accurately as possible. 

Probabilistic programming is a powerful method for programming nonparametric models. \emph{Stochastic memoization}~\cite{royStochasticProgrammingPerspective2008,woodStochasticMemoizerSequence2009} 
has been identified as a particularly useful technique in this. This paper is about semantic foundations for stochastic memoization.

In deterministic memoization~\cite{michieMemoFunctionsMachine1968}, the idea is to compute a function the first time it is called with a particular argument, and store the result in a memo-table. When the function is called again with the same argument, the memo-table is used, resulting in performance improvement but no semantic difference. Stochastic memoization is this memoization applied to functions that involve random choices, and so a memoized function is semantically different from a non-memoized one, because the random choices will only be made once for each argument.

We illustrate this with a simple example; this is informal and we consider a precise language and semantics in Section~\ref{sec:syntax}. Consider a function $f$ that returns a random number $[0,1]$ for each argument. It might be written $f(x) = \mlstinline{uniform}$. One run of the program might call $f$ with various arguments, and example runs are as follows:

\begin{center}\begin{tabular}{l|lllllll}
  \textit{Calls to $f$ in a particular run of a program}:&$f(0)$&$f(1)$&$f(0)$&$f(2)$&$f(1)$&$f(3)$&\dots\\
                                                           \hline
\textit{Results of calls in a run without memoization:}&0.43&0.01&0.72&0.26&0.48&0.16&\dots\\
  \textit{Results of calls in a run with memoization:}&0.43&0.01&\textbf{0.43}&0.26&\textbf{0.01}&0.16&\dots
\end{tabular}\end{center}
\smallskip
Thus in the memoized version, when the function is called again with the same value, the previous result is recalled, and the random choices are not made again. (Note that although this is called `stochastic memoization', the terminology is perhaps confusing: the memoization always happens, and it is not `randomly deciding whether or not to memoize'.)

From a semantic perspective, the role of stochastic memoization is clear when we use a monad-based interpretation with a probability monad \lstinline|Prob|.
This might be thought of as the Giry monad~\cite{giry} or a probabilistic powerdomain~\cite{jia-monad,jgl-domain-probprog}, or a Haskell monad (e.g.~\cite{lazyppl-popl}).

A distribution on a type \lstinline|b| with parameters from \lstinline|a| has type \lstinline|a -> Prob(b)|. On the other hand, a random function is a probability distribution on the type of deterministic functions, having type \lstinline|Prob(a -> b)|. Whereas parameterized distributions are a key idea in parametric statistics, random functions are a key idea in nonparametric statistics. And stochastic memoization is a higher-order function with probabilistic effects, of type 
\begin{lstlisting}[columns=flexible]
  mem :: (a -> Prob b) -> Prob (a -> b) 
\end{lstlisting} 
that converts parameterized distributions into random functions, by making the random choice once for each argument. 
This \lstinline|mem| combinator plays a crucial role in Church~\cite{goodmanChurchLanguageGenerative2008} and WebPPL~\cite{dippl}, and appears with this type in our Haskell library LazyPPL~\cite{LazyPPL}. Stochastic memoization also plays a role in Blog~\cite{blog}, Hansei~\cite{LogicProgrammingHANSEI}, and many other languages (e.g.~\cite{mathematica-stochmem,problog}). It is not difficult to implement stochastic memoization, by using a memo-table. Nonetheless, its semantic properties remain elusive and developers have noted bugs and complications (e.g.~\cite{webppl-bug,hansei-issue}). Moreover, the existing semantic models of probability (such as~\cite{qbs,jia-monad,jgl-domain-probprog}) only support \lstinline|mem| for very restricted domain types~\lstinline|a| (see \S\ref{sec:examples}). In particular our own Haskell library~\cite{LazyPPL} supports stochastic memoization but the recent semantic analysis~\cite{lazyppl-popl} only explains it at certain domain types. The point of this paper is to extend this semantic analysis of stochastic memoization to a broader class of domains.

\paragraph{First example: White noise in a non-parametric clustering model. }
One common first example of stochastic memoization is as follows. Suppose we have a finite set of individuals, and we want to group them into an unknown number of clusters, and then assign attributes to the clusters. For example, we may want to form clusters and consider attributes on the clusters such as `Brexit-supporters', `mean geographic latitude/longitude', `geographic variance', `mean salary', and so on. A popular route is the `Dirichlet process with memoization', as follows, for which a generative model has the following pseudocode (see e.g.~\cite{royStochasticProgrammingPerspective2008,probmods-non-parametric,dippl}\cite{gv-bnp}):
\begin{enumerate}
\item We randomly decide which proportion of individuals are in each cluster. We assign a unique identifier to each cluster, from some space~$\Atoms$ of identifiers. One might use the Dirichlet process with a diffuse base measure on $\Atoms$, for example the normal distribution on the real numbers.
\item Assign attributes to the cluster identifiers. For example, depending on whether that cluster supports Brexit, assign either true or false to the identifier. This particular assignment is a sample from a random function in $(\Atoms\to 2)$. This distribution might come from memoizing a constant Bernoulli distribution, assigning `true' to any cluster identifier with probability $0.5$.
\item
  Steps (i)-(ii) are generative, and we could run them to get some synthetic data. The idea of Bayesian clustering is to start with steps (i)--(ii) as a reasonable \emph{prior} distribution, in generative form, and to combine this with actual data to arrive at a \emph{posterior} distribution.
 In this example the actual data might come from a telephone survey, and we use conditional probability (aka Bayesian inversion) to arrive at a posterior distribution on the cluster proportions and their attributes. We can then use this to make predictions. The constant Bernoulli memoization is a reasonable prior for Brexit support, but the posterior will typically be much more complicated, with various correlations, etc.
\end{enumerate}
In this paper, we focus on step~(ii), stochastic memoization: steps~(i) and~(iii) are studied extensively elsewhere (e.g. see \cite{gv-bnp} in the statistics literature, or~\cite{dirichlet-is-natural,staton:sfinite,probprogbook} in the semantics literature, and references therein).

This simple example of a memoized constant Bernoulli function is easy to implement using a memo-table, but already semantically complicated. If we put $\Atoms=\mathbb{R}$, the real numbers, for the base measure, as is common in statistical modelling, then the memoized constant Bernoulli distribution on $(\Atoms\to 2)$ is 1-dimensional white noise: intuitively, for every $x\in\mathbb{R}$ we toss a coin to pick true or false, making an uncountable number of independent random choices. (As an aside, we note that we could combine steps~(i) and~(ii), using a complicated base measure for the Dirichlet process that includes all the attributes. This model would not be compositional, and in any case, some kind of memoization would still be needed to implement the Dirichlet process.)
\vspace{-.2in}
\paragraph{Challenge.}
In this paper, we address the challenge of showing that the following items are consistent:
\begin{enumerate}
\item[(1)] a type $\Atoms$ with a diffuse probability distribution (Def~\ref{eqn:diffuse});
\item[(2)] a type $\tbool$ of Booleans with Bernoulli probability distributions (i.e.~tossing coins, including biased coins);
\item[(3)] a type of functions $[\Atoms\to \tbool]$, with function application (\ref{eqn:fn-app});
\item[(4)] stochastic memoization of the constant Bernoulli functions \eqref{eqn:mem-const-bernoulli};
\item[(5)] the language supports the dataflow property (Def.~\ref{eqn:dataflow}).
\end{enumerate}
These items are together inconsistent with traditional measure theory, as we discuss in Section~\ref{sec:mem-nonatomic}, where we also make the criteria precise. Nonetheless (1)-(4) are together easy to implement in a probabilistic programming language, and useful for Bayesian modelling. Item~(5) is a very useful property for program reasoning and program optimization. Item~(5) is also a fundamental conceptual aspect of axiomatic probability theory, since in the measure-theoretic setting it amounts to Fubini's theorem~\cite{kock} and the fact that probability measures have mass $1$, and in the categorical abstraction of Markov categories~\cite{fritz} it amounts to the interchange law of affine monoidal categories.

There \emph{are} measure-theoretic models where some of these items are relaxed (\S\ref{sec:fin-mem}--\ref{sec:mem-nonatomic}). For example, if we drop the requirement of a diffuse distribution, then there are models using Kolmogorov extension (\S\ref{sec:countable}).

A grand challenge is to further generalize these items, for example to allow memoization of functions $A\to B$ for yet more general~$A$ and~$B$, and to allow memoization of all definable expressions. Since the above five items already represent a significant challenge, and our semantic model is already quite complicated, we chose to focus on a `minimal working example' for this paper. 

To keep things simple and minimal, in this paper we side-step measure-theoretic issues by noticing that the equations satisfied by a diffuse probability distribution are exactly the equations satisfied by name generation (e.g.~\cite[\S VB]{staton-instances}). Because of this, we can use categorical models for name generation (following e.g.~\cite[\S4.1.4]{moggi-techreport}, \cite[\S3.5]{stark-thesis}) instead of traditional measure theory. Name generation can certainly be implemented using randomness, and there are no clashes of fresh names if and only if the names come from a diffuse distribution (see also e.g.~\cite{nu-calculus}). On the other hand, if we keep things simple by regarding the generated names as \emph{pure names}~\cite{milner-names}, we avoid any other aspects of measure theory, such as complicated manipulations of the real numbers. 

\paragraph{Contributions.}
To address the challenge of the consistency of items~(1)--(5) above, our main contributions are then as follows.

\begin{enumerate}
\item We first provide an operational semantics for a minimal toy probabilistic programming language that supports stochastic memoization and name generation (\S\ref{sec:operational}).

\item We then (\S\ref{sec:denotational-semantics}) construct a cartesian closed (for function spaces) categorical model of this language endowed with an affine commutative monad (Theorem~\ref{thm:probabilistic-local-state-monad-commutative-affine}). In common with other work on local state~(e.g.~\cite{plotkinNotionsComputationDetermine2002,klms-ground-state-monad}), we use a functor category semantics, indexing sets by possible worlds. In this paper, those worlds are finite fragments of a memo-table. 

\item We prove that our denotational semantics is sound with respect to the operational semantics, ensuring the correctness of our approach and validating that lines can be reordered in the operational semantics (Theorem~\ref{thm:soundness}). The class of functions that can be memoized includes constant Bernoulli functions. We call these functions \emph{freshness-invariant} (Definition~\ref{def:freshness-invariant}).

  The soundness theorem (\ref{thm:soundness}) is not trivial because the timing of the random choices differs between the operational and denotational semantics. In the operational semantics, the memo-table is partial, and populated lazily as needed, when functions are called with arguments. This is what happens in all implementations. However, this timing is intensional, and so by contrast, in the denotational semantics, the memo-table is always totally populated as soon as the current world is extended with any functions or arguments. 

\item Finally, we present a practical Haskell implementation~\cite{stoch-mem-implementation-github} which compares the small-step, big-step operational, and denotational semantics, demonstrating the applicability of our results (\S\ref{sec:implementation}).

\end{enumerate}




\section{Stochastic memoization by example}
\label{sec:background}

This section discusses the law of stochastic memoization and provides examples in finite, countable, and non-enumerable domain settings. We then address the challenges posed by the naive use of the state monad, and we clarify our objective: finding a model of probability that supports stochastic memoization over non-enumerable domains, satisfying the dataflow property, and that has function spaces.

In what follows, we use two calculi: (a) The internal metalanguage of a cartesian closed category with a strong monad \lstinline|Prob|, for which we use Haskell notation, but which is roughly Moggi's monadic metalanguage~\cite[\S2.2]{moggi:computation_and_monads}. (b)~An ML-like programming language which is more useful for practical programming, but which would translate into language~(a); this is roughly Moggi's `simple programming language'~\cite[\S2.3]{moggi:computation_and_monads}. We assume passing familiarity with probability and monadic programming in this section, but the informal discussion here sets the context, and we move to more formal arguments in Section~\ref{sec:syntax}.

(Recall some Haskell notation: we write \lstinline|\x -> t| for lambda abstraction; \lstinline|>>=| for monadic bind, i.e.~Kleisli composition; \lstinline|return| for the unit; a \lstinline|do| block allows a sequence of monadic bound instructions. We write \lstinline|const x| for the constant \lstinline|x| function, \lstinline|const x = \y -> x|.) 

\paragraph{Memoization law.}
\begin{definition}\label{def:mem}
  A strong monad \emph{supports stochastic memoization of type }\lstinline|a->b| if it is equipped with a morphism
  \lstinline|mem :: (a -> Prob b) -> Prob (a -> b)| that satisfies the following equation in the metalanguage, for every \lstinline[mathescape]|x$\cb_0$ :: a| and
  \lstinline|f :: a -> Prob b|:
  \begin{equation}\label{eqn:mem-monad}
    \mlstinline{mem f     =    f x$_0\,$ >>= (\\y$_0\,$ -> mem f >>= (\\fMem -> return (\\x -> if x == x$_0\,$ then y$_0\,$ else fMem x) ) )}
  \end{equation}
\end{definition}

As noted at the beginning of this section, we will pass between an internal metalanguage for strong monads, and an ML-like programming language that would be interpreted using strong monads. In Section~\ref{sec:syntax} we introduce this programming language precisely, but for now we note that it has a special syntax $\lambdamem{x} u$, meaning \lstinline[mathescape]|mem (\$x$ -> $u$)|, since this is a common idiom\footnote{borrowing Melliès' use of the Hebrew letter $\mathsf{\mem}$ (``mem'')~\cite{melliesLocalStatesString2014}}. The law of Definition~\ref{def:mem} requires equations such as: 
\begin{equation}\label{eqn:stochmem}\begin{minipage}[!htb]{.4\textwidth}
\begin{minipage}[!htb]{.3\textwidth}
   \[\begin{aligned}&\iletin f {\lambdamem{{ x}}{{u}}\\&}{f@ n}\end{aligned}
\qquad{\large\overset{\mathclap{\text{1 sample}}}{=}}\qquad
u[n/x]\]
\end{minipage}

\end{minipage}
\begin{minipage}[!htb]{.55\textwidth}
\begin{minipage}[!htb]{.3\textwidth}
  \[\begin{aligned}
      &\iletin f{\lambdamem{x}{u}}
{\\&\iletin {v_1} {f @ n }{
      \\&\iletin {v_2}{f@ n}{
\\&\ireturn (v_1, v_2)}}}\end{aligned}
\qquad{\large \overset{\mathclap{\text{several samples}}}{=}}\qquad\quad
\begin{aligned}
  &\iletin v {u[n/x]}{\\
&\ireturn (v, v)}\end{aligned}\]
\end{minipage}
\end{minipage}\end{equation}

\label{sec:examples}

The examples in the introduction use memoization of a constant Bernoulli function, i.e.
\begin{equation}\label{eqn:mem-const-bernoulli}
  \mlstinline{mem (\\x ->bernoulli p)  =  mem (const (bernoulli p)) :: Prob(a -> Bool)}
\end{equation}
i.e.~$\lambdamem x\mlstinline{bernoulli p}$, 
where \lstinline|bernoulli p :: Prob Bool| is a Bernoulli probability distribution (biased coin toss) with bias~\lstinline|p|. An intuition is that this is a binary white noise; every point in~\lstinline|a| has an independently chosen random Boolean value. 

Notice that for the laws we have also needed function application
\begin{equation}\label{eqn:fn-app}
  @ \quad\mlstinline{   :: ((a -> b) , a) -> b}
  \end{equation}
In summary, memoized constant Bernoulli functions~\eqref{eqn:mem-const-bernoulli}, and function application~\eqref{eqn:fn-app}, are a bare minimum to discuss semantic issues around stochastic memoization. 

We now consider interpretations where the domain \lstinline|a| is finite (\S\ref{sec:fin-mem}), countable (\S\ref{sec:countable}), and uncountable (\S\ref{sec:mem-nonatomic}).

\subsection{Memoization with finite domain}
\label{sec:fin-mem}

For finite domains \lstinline|a|, memoization is straightforward. It involves simply sampling a value of $f(x)$ for all inhabitants of $x ∈ X$ and returning the assignment as a finite mapping. For example, when $\mlstinline{a} = \mlstinline{bool}$, we can implement memoization in Haskell as follows:
\begin{lstlisting}[language=haskell]
   mem f   =   do { fT <- f True ; fF <- f False ; return (\b -> if b then fT else fF)}
 \end{lstlisting}

\paragraph{Semantic interpretation with finite domain.} Memoization with finite domains is supported by a denotational semantics using any strong monad. For example, the category of sets and the monad of finitely supported probability distributions (e.g.~\cite{jacobs-probabilities}). For $\mlstinline a=\mlstinline{bool}$, this is nothing but the double-strength:
\[
  \mlstinline{(bool -> Prob b)} \cong \mlstinline{(Prob b,Prob b)} \xrightarrow{\text{double-strength}} \mlstinline{Prob (b,b)}\text.
\]
For other finite $\mlstinline{a}$, it is defined using the double-strength by induction.

\subsection{Memoization with countable/enumerable domain}
\label{sec:countable}
When \lstinline|a| is enumerable, such as \lstinline|a=Int|, memoization is useful for defining point processes. Memoization can be regarded as providing an infinite stream of random choices, since the streams over \lstinline|b| are isomorphic with the functions
\lstinline|a -> b|. 

Infinite streams of random choices are crucial examples of statistical processes~\cite{gv-bnp}. For an example of an application, recall the one-dimensional Poisson point process. This is a random sequence of real numbers in which the gaps between consecutive numbers are exponentially distributed. Assuming an exponential distribution with a given rate, \lstinline|exponential rate :: Prob RealNum|, we can sample a sequence of these exponential gaps from \lstinline|mem (const (exponential rate)) :: Prob (Int -> RealNum)|.
To get the corresponding list of points of the Poisson point process (with exponential interoccurence times), we simply keep a cumulative sum total of the points, starting from the \lstinline|lower| point:

\begin{lstlisting}[language=haskell]
   poissonPP :: Double -> Double -> Prob [Double]
   poissonPP lower rate = do { gaps <- mem (const (exponential rate)) ; return (scanl1 (+) lower (map gaps [1 .. ])) }
\end{lstlisting}

We implement memoization with enumerable \lstinline|a| in the Haskell LazyPPL library~\cite{lazyppl-popl} without using state, instead using Haskell's laziness and tries, following~\cite{hinzeGeneralizingGeneralizedTries2000} (see~\cite{lazyppl-popl}). We use the Poisson process extensively in the demonstrations for LazyPPL~\cite{LazyPPL}. 



\paragraph{Semantic interpretation with enumerable domains.} Memoization with enumerable domains is supported by a denotational semantics using the category of measurable spaces and the Giry monad~\cite{giry}. Although the category is not Cartesian closed, the function space $B^\NN$ \emph{does} exist for all standard Borel $B$, and is given by the countable product of $B$ with itself, $\prod_\NN B$. Memoization amounts to using Kolmogorov's extension theorem to define a map $(G\,B)^\NN\to G(B^\NN)$ (see \cite[\S4.8]{pollard-book} and~\cite[Thm.~2.5]{dirichlet-is-natural}).

\hide{\paragraph{Random graphs} Stochastic memoization is particularly useful for programming statistical models with random relational structures, such as graph social networks, the world wide web, or biochemical pathways. When modeling relational data, one typically represents observations of binary relationships between object collections with random arrays $(X_{i, j})_{i, j}$ of Boolean random variables, where $i$ and $j$ range over countable collections of potentially related objects.

When comparing a single countable object collection pairwise (i.e., $i$ and $j$ range over the same countable index), the array $(X_{i, j})_{i, j}$ can be viewed as the adjacency matrix of a random graph. These random adjacency matrices $(X{i, j})_{i, j}$ are considered \emph{exchangeable} if their joint distribution remains invariant under node relabeling in the corresponding graph. Assuming the random graph is simple (i.e., $(X{i, j})_{i, j}$ is symmetric with zero diagonal), the Aldous-Hoover theorem posits that every associated exchangeable adjacency matrix can be parameterized by a random measurable function $G: [0, 1]^2 ⟶ [0, 1]$ called as a \emph{graphon}. For two nodes $i$ and $j$ in the random simple graph, $G(u_i, u_j)$ is the edge probability between them, with $u_i$ and $u_j$ being independent and identically distributed (iid) random variables. Consequently, Bayesian models for exchangeable simple graphs are fully determined (in distribution) by a prior over graphons.

Consider, for instance, a Haskell typeclass \lstinline|RandomGraph|, corresponding to an abstract type whose interface enables generating a new random graph (with \lstinline|newGraph|) from a seed \lstinline|g|, drawing vertices at random (with \lstinline|newVertex|), and examining the presence of edges between vertices (with \lstinline|isEdge|). In line with the Aldous-Hoover theorem~\cite{aldousRepresentationsPartiallyExchangeable1981,hooverRelationsProbabilitySpaces1979,orbanz-roy}, every measurable function $\mlstinline{g}: [0, 1]^2 → [0, 1]$ (a \emph{graphon}) can serve as a seed, resulting in the following implementation in the LazyPPL library:

\begin{minipage}[!htb]{.51\textwidth}
\begin{lstlisting}[language=haskell]
class RandomGraph g where
	type Graph g
	data Vertex g
	newGraph:: g -> Prob (Graph g)
	newVertex:: Graph g -> Prob (Vertex g)
	isEdge:: Graph g -> Vertex g -> Vertex g -> Bool
newtype Graphon = G ((*'$\mathbb{R}$'*, *'$\mathbb{R}$'*) -> *'$\mathbb{R}$'*)
\end{lstlisting}
\end{minipage}
\,
\begin{minipage}[!htb]{.49\textwidth}
\begin{lstlisting}[language=haskell]
instance RandomGraph Graphon where
	type Graph Graphon = (*'$\mathbb{R}$'*, *'$\mathbb{R}$'*) -> Bool
	data Vertex Graphon = V *'$\mathbb{R}$'*
	-- return a random function '(*'$\color{gray}\mathbb{R}$'*, *'$\color{gray}\mathbb{R}$'*) -> Bool'
	newGraph (G graphon) = mem $ bernoulli . graphon
	newVertex _ = V <$> uniform
	isEdge graph (V x) (V y) = graph (x, y)
\end{lstlisting}
\end{minipage}

The memoization function \lstinline[mathescape]|mem :: (($ℝ$, $ℝ$) -> Prob Bool) -> Prob (($ℝ$, $ℝ$) -> Bool)| ensures that once an edge between $x$ and $y$ is sampled (with probability \lstinline|graphon (x, y)|), its presence or absence remains constant throughout the rest of the program execution. For example, \lstinline|newGraph (G (const 0.5))| generates the Erdős-Rényi~\cite{erdosRandomGraphs1959,radoUniversalGraphsUniversal1964,ackermannWiderspruchsfreiheitAllgemeinenMengenlehre1937} random graph.}

\subsection{Memoization with non-enumerable/diffuse domain }
\label{sec:mem-nonatomic}
We now move beyond enumerable domains, to formalize the challenge from Section~\ref{sec:intro}.
In Section~\ref{sec:intro} we illustrated this with a clustering model. See~\cite{LazyPPL} for the full implementation in our Haskell library, LazyPPL, along with other models that also use memoization, including a feature extraction model that uses the Indian Buffet Process, and relational inference with the infinite relational model (following~\cite{probmods-non-parametric}).

Rather than axiomatizing uncountability, we consider diffuse distributions. 
\begin{definition}[Diffuse distribution]\label{eqn:diffuse}
  Let \lstinline|a| be an object with an equality predicate (\lstinline|(a,a)->bool|). A \emph{diffuse distribution}\footnote{Diffuse measures are often called `atomless' in probability theory. We will also want to regard names in name generation as atomic, so we avoid this clash of terminology.} is a term \lstinline|p| such that 
  \[\mlstinline{do \{x <- p ; y <- p ; return (x ==y)\}}
    \qquad\text{is semantically equal to }\qquad
    \mlstinline{return (false)}\text.
  \]
  \end{definition}
  For example, in a probabilistic programming language over the real numbers, we can let \lstinline|a| be the type of real numbers and let \lstinline|p| be a uniform distribution on $[0,1]$, or a normal distribution, or an exponential distribution. These are all diffuse in the above sense. The Bernoulli distribution on the booleans is not diffuse, because there is always a chance that we may get the same result twice in succession. 

  For the reader familiar with traditional measure theory, we recall that if \lstinline|p| is diffuse then \lstinline|a| is necessarily an uncountable space. For any probability distribution on a countable discrete space must give non-zero measure to at least one singleton set.
  
The implementation trick using tries from Section~\ref{sec:countable} will not work for diffuse measures, because we cannot enumerate the domain of a diffuse distribution. It is still possible to implement memoization using state and a memo-table (e.g.~\cite{LazyPPL}). Unlike a fully stateful effect, however, in this paper we argue that stochastic memoization is still compatible with commutativity/dataflow program transformations:

\begin{definition}[Dataflow property]\label{eqn:dataflow}
	A programming language is said to have the \emph{dataflow property} if program lines can be reordered (commutativity) and discarded (discardability, or affineness) provided that the dataflow is preserved.
	In other words, the language satisfies the following commutativity and discardability equations: 

  \begin{align}
    \hspace{-3mm}\mlstinline{do \{x1 <- t1 ; x2 <- t2 ; u\}}\ &=\ \mlstinline{do \{x2 <- t2 ; x1 <- t1 ; u\}}
      \label{eq:let-comm}\\
    \mlstinline {do \{x1 <- t1 ; t2\}} & = \mlstinline{t2}
                                         \label{eq:let-affine}
                                         &\text{where $\mlstinline{x1} ∉ \fv(\mlstinline{t2})$ and $\mlstinline{x2} ∉ \fv(\mlstinline{t1})$.} 
  \end{align}
\end{definition}
The dataflow property expresses the fact that, to give a meaning to programs, the only thing that matters is the topology of dataflow diagrams. 
These transformations are very useful for inference algorithms and program optimization. But above all, on the foundational side, dataflow is a fundamental concept that corresponds to monoidal categories and is crucial to have a model of probability. As for monoidal categories, a strong monad is commutative \eqref{eq:let-comm} if and only if its Kleisli category is monoidal (commutativity is the monoidal interchange law), and affine \eqref{eq:let-affine} if the monoidal unit is terminal. In synthetic probability theory, dataflow is regarded by various authors as a fundamental aspect of the abstract axiomatization of probability:
Kock~\cite{kock-comm-monad} argues that any monad that is strong commutative and affine can be abstractly viewed as a probability monad, and affine monoidal categories are used as a basic setting for synthetic probability by several authors~\cite{fritz,cho-jacobs,dario-thesis,dario}.
The reader familiar with measure-theoretic probability will recall that the proof that the Giry monad satisfies~(\ref{eq:let-comm}) amounts to Fubini's theorem for reordering integrals (e.g.~\cite{staton:sfinite}).



\paragraph{Semantic interpretations for diffuse domains}\newcommand{\RR}{\mathbb{R}}
The point of this paper is to provide the first semantic interpretation for memoization of the constant Bernoulli functions~(\ref{eqn:mem-const-bernoulli}) with diffuse domain (Def.~\ref{eqn:diffuse}). 
We emphasize that although other models can support some aspects of this, there is no prior work that supports everything.
\begin{itemize}
\item With countable domain, there is a model in measurable spaces, as discussed in Section~\ref{sec:countable}. But there can be no diffuse distribution on a countable space.
\item In measurable spaces, we can form the uncountable product space~$\prod_\RR 2$ of $\RR$-many copies of~$2$. We can then define a white noise probability measure on $\prod_\RR 2$ via Kolmogorov extension (e.g.~\cite[4.9(31)]{pollard-book}). Moreover, there are diffuse distributions on $\RR$, such as the uniform distribution on $[0,1]$. However, it is known that there is no measurable evaluation map $\RR\times (\prod_\RR 2)\to 2$ (see~\cite{aumann}), and so we cannot interpret function application~(\ref{eqn:fn-app}).
\item In quasi-Borel spaces~\cite{qbs}, there is a quasi-Borel space $[\RR\to 2]$ of measurable functions, and a measurable evaluation map $\RR\times ([\RR\to 2)\to 2$, but there is no white noise probability measure on $[\RR\to 2]$. The intuitive reason is that, in quasi-Borel spaces, a probability measure on $[\RR\to 2]$ is given by a random element, i.e.~a morphism $\Omega\to[\RR\to 2]$, which curries to a measurable function $\Omega\times\RR\to 2$. But there is no such measurable function representing white noise (e.g.~\cite[Ex~1.2.5]{kallianpur}).
\item There are domain-theoretic treatments of probability theory that support Kolmogorov extension, uniform distributions on $\RR$, and function spaces~\cite{jia-monad,jgl-domain-probprog}. However, these treatments regard the real numbers $\RR$ as constructive, and hence there are no non-trivial continuous morphisms $\RR\to 2$, and there is no equality test on $\RR$, so that we cannot regard $\RR$ with a diffuse distribution as formalized equationally in Definition~\ref{eqn:diffuse}. The same concern seems to apply to recent approaches using metric monads\cite{mpp-wasserstein-alg}.
\item The semantic model of beta-bernoulli in~\cite{staton:betabernoulli} is a combinatorial model that includes aspects of the beta distribution, which is diffuse in measure theory. That model does not support stochastic memoization, but as a presheaf-based model it is a starting point for the model in this paper. 
\item There is a straightforward implementation of stochastic memoization that uses local state, as long as the domain supports equality testing \cite{LazyPPL}. The informal idea is to make the random choices as they are needed, and remember them in a memo-table, and keep this memo-table in a local state associated with the function. Therefore one could use a semantic treatment of local state to analyze memoization. For example, one could build a state monad in quasi-Borel spaces. However, state effects in general do not support the dataflow property (Def.~\ref{eqn:dataflow}), since we cannot reorder memory assignments in general. Ideally, one could use a program logic to prove that this particular use of state does support the dataflow property. Although there are powerful program logics for local state and probability (e.g.~\cite{bizjak-birkedal}), we have not been able to use them to prove this.
\end{itemize}
There are other models of higher-order probability (e.g.~\cite{measurable-cones,crubille-cones,cp-games-prob}). These do not necessarily fit into the monad-based paradigm, but there may be other ways to use them to address the core challenge in Section~\ref{sec:intro}.

\section{A language for stochastic memoization and name generation}
\label{sec:syntax}

Our probabilistic programming language has a minimal syntax, emphasizing the following key features:
\begin{itemize}
  \item \textbf{name generation}: we can generate fresh names (referred to as \emph{atomic} names or \emph{atoms}, in the sense of Pitts' nominal set theory~\cite{pittsNominalSetsNames2013}) with constructs such as $\letin{x}{\ifresh}{⋯}$. In the terminology of Def.~\ref{eqn:diffuse}, this is like a generic diffuse probability measure, since fresh names are distinct. 
  \item basic \textbf{probabilistic effects}: for illustrative purposes, the only distribution we consider, as a first step, is the Bernoulli distribution (but it can easily be extended to other discrete distributions). Constructs like $\letin{b}{\iflip}{⋯}$ amount to flipping a coin with bias $θ$ and storing its result in a variable $b$. 
  \item \textbf{stochastic memoization}: if a probabilistic function $f$ -- defined with the new $\lambdamem{}{}$ operator -- is called twice on the same argument, it should return the same result (\cref{eqn:stochmem}).
\end{itemize}

We have the following base types: $\tbool$ (booleans), $\Atoms$ (atomic names), and $\Fns$ (which can be thought of as the type of memoized functions $\Atoms\to \tbool$). For the sake of simplicity, we do not have arbitrary function types. In fine-grained call-by-value fashion~\cite{levy:cbpv}, there are two kinds of judgments: typed values, and typed computations. The grammar and typing rules of our language are given in Figure~\ref{fig:language}. The typing rules are standard, except for the $\lambdamem{}{}$ operator, which is the key novelty of our language. The typing rule for $\lambdamem{}{}$ is given in Figure~\ref{fig:language} and is explained in the next section. (Also, equality $v = w$ and memoized function application $v @ w$ are pure computations, \ie in the categorical semantics (\cref{sec:cat-semantics}), they will be composed by the unit of the monad.)

\begin{longtable}{|l c l|}
  \caption{Grammar and typing rules of the language} \label{fig:language}\\
  \hline
  \multicolumn{3}{|c|}{\textbf{Types}} \\


  \hline
  \endfirsthead
  \hline
  \endhead
  \hline
  \endfoot
  \endlastfoot

  $A, B$ & \mbox{::=} & \; $\tbool$ \mmid $\Atoms$ \mmid $\Fns$ \mmid $A × B$ \\ \hline \hline
  \multicolumn{3}{|c|}{\textbf{Expressions}} \\ 
  \hline
  \multicolumn{3}{|l|}{\textbf{Values:}} \\[-.5em]
  $v, w$ & \mbox{::=} & \; $\itrue$ \mmid $\ifalse$ \mmid $x$ \mmid $(v,w)$ \\
  \multicolumn{3}{|l|}{\textbf{Computations:}} \\[-.5em]
  $u, t$ & \mbox{::=} & \; $\ireturn(v)$ \mmid $\iletin x u t$ \mmid $\ite v u t$ \mmid $\imatch v {(x,y)} t$ \\
  & & \mmid $\iflip$ \mmid $\ifresh$ \mmid $v=w$ \mmid $\lambdamem x u$ \mmid $v@w$ \\[.5em] \hline \hline

  \multicolumn{3}{|c|}{\textbf{Typing judgements}} \\ \hline
  \multicolumn{3}{|l|}{\textbf{Typed values:}} \\
  \multicolumn{3}{|c|}{
    \begin{prooftree}  
      \hypo{-}
      \infer1{\vj{Γ}\itrue \tbool}
    \end{prooftree}
    \qquad
    \begin{prooftree}  
      \hypo{-}
      \infer1{\vj{Γ}\ifalse \tbool}
    \end{prooftree}
    \qquad
    \begin{prooftree}
      \hypo{-}
      \infer1{\vj{Γ,x:A, Γ'}x A}
    \end{prooftree}
    \qquad
    \begin{prooftree}  
      \hypo{\vj{Γ}v A} 
      \hypo{\vj{Γ}w B}
      \infer2{\vj{Γ}{(v,w)} {A× B}}
    \end{prooftree}}\\[1em]
\multicolumn{3}{|l|}{\textbf
  {Typed computations:}} \\[.5em]
  \multicolumn{3}{|c|}{
    \begin{prooftree}
      \hypo{\vj{Γ}v A}
      \infer1{\cj{Γ}{\ireturn(v)} A}
    \end{prooftree}
    \qquad
    \begin{prooftree}
      \hypo{\cj{Γ}u A}
        \hypo{\cj{Γ,x:A}t B}
        \infer2{\cj{Γ}{\iletin x u t} B}
        \end{prooftree}
        }\\[1em]
    \multicolumn{3}{|c|}{
        \begin{prooftree}
        \hypo{\vj{Γ}v \tbool}
        \hypo{\cj{Γ}u A}
        \hypo{\cj{Γ}t A}
        \infer3{\cj{Γ}{\ite v u t} A}
        \end{prooftree}
        \qquad
        \begin{prooftree}
        \hypo{\vj{Γ}v {A× B}}
        \hypo{\cj{Γ,x:A,y:B}t C}
        \infer2{\cj{Γ}{\imatch v {(x,y)} t} C}
        \end{prooftree}
        }\\[1em]
    \multicolumn{3}{|c|}{
        \begin{prooftree}
        \hypo{-}
        \infer1{\cj{Γ}{\iflip} \tbool}
        \end{prooftree}
        \qquad
        \begin{prooftree}
        \hypo{-}
        \infer1{\cj{Γ}{\ifresh} \Atoms}
        \end{prooftree}
        \qquad
        \begin{prooftree}
        \hypo{\vj{Γ}v \Atoms}
        \hypo{\vj{Γ}w \Atoms}
        \infer2{\cj{Γ}{(v=w)} \tbool}
        \end{prooftree}
        }\\[1em]
    \multicolumn{3}{|c|}{
        \begin{prooftree}
        \hypo{\cj{Γ,x:\Atoms}u \tbool}
        \infer1{\cj{Γ}{\lambdamem x u} \Fns}
        \end{prooftree}
        \qquad
        \begin{prooftree}
        \hypo{\vj{Γ}v \Fns}
        \hypo{\vj{Γ}w \Atoms}
        \infer2{\cj{Γ}{(v@w)} \tbool}
        \end{prooftree}
        }\\[1em]
  \hline
\end{longtable}

\section{Operational Semantics}
\label{sec:operational}

We now present a small-step operational semantics for our language.  The operational semantics defines the rules for reducing program expressions, which form the basis for understanding the behavior of programs written in the language. Henceforth, we fix a countable set of variables $x, y, z, … ∈ \Var$, and consider the terms up to $α$-equivalence for the $\lambdamem{}{}$ operator. Since we focus on functions with boolean codomain, our partial memo-tables are represented as partial bigraphs (bipartite graphs). 

\begin{definition}[Partial bigraph]\label{def:bigraph}
  A partial bigraph $𝖌 ≝ (𝖌_L, 𝖌_R, E)$ is a finite bipartite graph where the edge relation $E：𝖌_L × 𝖌_R → \{\itrue, \ifalse, ⊥\}$ is either true, false or undefined ($⊥$) on each pair of left and right nodes $(ƒ, a) ∈ 𝖌_L × 𝖌_R$. In the following, left nodes will be thought of as function labels and right nodes as atom labels.
  By abuse of notation, syntactic truth values will be conflated with semantic ones. For a partial graph $𝖌$, $E(ƒ, a) = β ∈ \{\itrue, \ifalse, ⊥\}$ will be written $ƒ \xto {β} a$ when $𝖌$ is clear from the context. 
\end{definition}

\subsection{Extended expressions}

We introduce extended expressions $e$, by extending the grammar of computations (\ref{fig:language}) with an extra construct $\memobraces{u}$, where $u$ is a computation,$(ƒ, a)$ is a pair of function and atom labels to memoize, and $γ$ is the environment to restore after the result of $ƒ$ at $a$ has been computed and stored. 
Intuitively, the decoration $\memobraces{-}$ is thought of as a memoization context, indicating expressions where memoization should happen: $\memobraces{u}$ is a computation that memoizes the result of $u$, and then restores the environment to the state it was in before the computation $u$ was evaluated.
In the following, $Δ ∈ \bigcup_{n ≥ 0} (𝖌_L × 𝖌_R)^n$ is a finite stack of function--atom label pairs, indicating that we are in the process of computing the result of these functions at these atoms for the first time. Each newly introduced function--atom label pair is assumed not to already belong to the memoization stack.

\begin{longtable}{|c c|}
  \caption{Extended expression typing rules. }\label{fig:extended_expressions}\\
  \hline
  \multicolumn{2}{|c|}{\textbf{Extended expression typing judgements.} Here, $(ƒ, a) ∉ Δ ∪ Δ_1 ∪ Δ_2$.}\\
  
    \hline
    \multicolumn{2}{|c|}{} \\[-2ex]
    \endfirsthead
    \hline
    \multicolumn{2}{|c|}{} \\[-2ex]
    \endhead
    \hline
    \endfoot
    \hline
    \endlastfoot

  \begin{prooftree}
      \hypo{\cj{Γ} u A}
      \infer1{\cje{Γ} {∅} u A}
  \end{prooftree}
  \qquad
  \begin{prooftree}
      \hypo{\cje{Γ} {Δ} u A}
      \infer1{\cje{Γ} {(ƒ, a), Δ} {\memobraces{u}} A}
    \end{prooftree}
    &
    \begin{prooftree}
      \hypo{\cje {Γ} {Δ_1} u A}
      \hypo{\cje {Γ,x:A} {Δ_2} t B}
      \infer2{\cje {Γ} {Δ_1, Δ_2} {\iletin x u t} B}
  \end{prooftree}
  \\[1.5em]
  \begin{prooftree}
      \hypo{\cje {Γ} {Δ_1} u A}
      \hypo{\cje {Γ,x:A} {Δ_2} t B}
      \infer2{\cje {Γ} {(ƒ, a), Δ_1, Δ_2} {\iletin x {\memobraces{u}} t} B}
    \end{prooftree}
    &
  \begin{prooftree}
      \hypo{\cje {Γ} {Δ_1} u A}
      \hypo{\cje {Γ,x:A} {Δ_2} t B}
      \infer2{\cje {Γ} {(ƒ, a), Δ_1, Δ_2} {\iletin x u {\memobraces{t}}} B}
  \end{prooftree}\\[1.5em]
  \multicolumn{2}{|c|}{
    \begin{prooftree}
      \hypo{\vj{Γ}{v}{\tbool}}
      \hypo{\cje{Γ}{Δ_1}{u}{A}} 
      \hypo{\cje{Γ}{Δ_2}{t}{A}}
      \infer3{\cje{Γ}{Δ_1,Δ_2}{\ite v u t} A}
  \end{prooftree}}\\[1.5em]
  \begin{prooftree}
      \hypo{\vj{Γ}{v}{\tbool}}
      \hypo{\cje{Γ}{Δ_1}{u}{A}} 
      \hypo{\cje{Γ}{Δ_2}{t}{A}}
      \infer3{\cje{Γ}{(ƒ, a),Δ_1,Δ_2}{\ite v {\memobraces{u}} t} A}
  \end{prooftree}
    &
  \begin{prooftree}
      \hypo{\vj{Γ}{v}{\tbool}}
      \hypo{\cje{Γ}{Δ_1}{u}{A}} 
      \hypo{\cje{Γ}{Δ_2}{t}{A}}
      \infer3{\cje{Γ}{(ƒ, a),Δ_1,Δ_2}{\ite v u {\memobraces{t}}} A}
  \end{prooftree}\\[1.5em]
  \begin{prooftree}
      \hypo{\vj{Γ}{v}{A × B}}
      \hypo{\cje{Γ,x:A,y:B}{Δ}{t}{C}}
      \infer2{\cje{Γ}{Δ}{\imatch v {(x,y)} t} C}
  \end{prooftree}
    &
  \begin{prooftree}
      \hypo{\vj{Γ}{v}{A × B}}
      \hypo{\cje{Γ,x:A,y:B}{Δ}{t}{C}}
      \infer2{\cje{Γ}{(ƒ, a), Δ}{\imatch v {(x,y)} {\memobraces{t}}} C}
  \end{prooftree}\\[1.5em]
\end{longtable}

\subsection{Configurations}

We now define the set-theoretic interpretation of contexts. Context values are built by combining booleans, atomic names and functions using pairing. Thus a context value is a tree, where the branches are understood as pairing. 

\begin{definition}
  If $S$ is a finite set, $\Tree[S] ≅ \biguplus_{ n ≥ 0} C_n \, S^{n+1}$ (where $C_n$ is the $n$-th Catalan number, and $C_n \, S^{n+1}$ is a coproduct of $n$ copies of $S^{n+1}$, one for each possible bracketing) denotes the set of all possible non-empty trees with internal nodes the cartesian product and leaf nodes taken in $S$.
  \begin{example}
  If $S ≝ \{s_1, s_2\}$, then $s_1 ∈ \Tree[S], (s_2, s_1) ∈ \Tree[S], (s_1, (s_1, s_2)) ∈ \Tree[S], …$
  \end{example}
\end{definition}

\begin{definition}[Set-theoretic denotation of contexts.]
  Let $𝖌$ be a partial bigraph. The set-theoretic denotation $\denpar { - }$ of a context $Γ$
  is defined as $\denpar {\tbool} ≝ 2 ≅ \{ \itrue, \, \ifalse\}, \, \denpar {\Fns} ≝ 𝖌_L, \, \denpar {\Atoms} ≝ 𝖌_R$ and $\denpar {-}$ is readily extended to every context $Γ$. Moreover, in the following, $γ ∈ \denpar {Γ} ⊆ \Tree^{\Var}$ denotes a context value.
\end{definition}

\begin{example} If $Γ ≝ (x: \tbool, y: \Fns, z: ((\Fns × 2) × \Atoms))$, then $\denpar {Γ} ≝ \{ x ↦ 2, y ↦ 𝖌_L, z ↦ ((𝖌_L × 2) × 𝖌_R)$ and an example of a context value is $γ ≝ \{ x ↦ \itrue, y ↦ ƒ_0, z ↦ ((ƒ_1, \itrue), a_0)\}$.
\end{example}

We now present terminal computations, redexes, reduction contexts, and configurations (\cref{fig:configurations}). Configurations encapsulate the computation state (a context value, an extended expression, a partial graph, and a map from the partial graph to closures), which helps keep track of different parts of the program as the computation proceeds.


\begin{longtable}{|l c l|}
  \caption{Terminal computations, redexes, reduction contexts, and configurations.}
  \label{fig:configurations}\\
\hline
\multicolumn{3}{|c|}{\textbf{Terminal computations $r$, Redexes $ρ$, and Reduction contexts $\C[-]$}} \\

\hline
\endfirsthead
\hline
\endhead
\hline
\endfoot
\endlastfoot

$r$ & \mbox{::=} & \; $\ireturn(v)$ \mmid $\lambdamem x u$ \mmid $\ifresh$ \\

$ρ$ & \mbox{::=} & \; $\iletin x r u$  \mmid $\memobraces{\ireturn (v)}$ \qquad where $ƒ ∈ 𝖌_L,\; a ∈ 𝖌_R,\; γ ∈ \Tree^{\Var}$\\
& & \mmid $\imatch v {(x,y)} t$ \mmid $\ite v t u$  
\mmid $\iflip$ \mmid $(v=w)$ \mmid $(v@w)$ \\

$\C[-]$ & \mbox{::=} & \; $[-]$ \mmid $\iletin x {\C[-]} u$ \mmid $\memobraces{\C[-]}$ \\[.3em] 
\hline

\multicolumn{3}{|c|}{\textbf{Configurations $(γ, u, 𝖌, λ)$}} \\ 
\cline{1-3} 
&\multicolumn{2}{l|}{$γ ∈ \Tree^\Var$ is a context value.} \\
&\multicolumn{2}{l|}{$u$ is an extended expression $\cje{Γ}{Δ}{u}{A}$.} \\
&\multicolumn{2}{l|}{$𝖌 ≝ (𝖌_L, 𝖌_R, E)$ is a partial graph.} \\
&\multicolumn{2}{l|}{$λ：𝖌_L → \Closures$, where $\Closures ≝ \big\{ (\lambdamem x u, γ) \;\mid\; \cj{Γ}{\lambdamem x u}{\Fns} \text{ and } γ ∈ \denpar{Γ}\big\}$}\\[.3em]
\hline
\end{longtable}

\subsection{Reduction rules}

Let $\denpar[γ]{-}$ be the function evaluating an expression value in a context value $γ$ (\eg $\denpar[γ]{x} = γ(x), \denpar[γ]{\itrue} = \itrue$). 

We can define the operational semantics of the language using reduction rules. They provide a step-by-step description of how expressions are evaluated and transformed during execution, following a left-most outer-most strategy, with lexical binding. Given a configuration $(γ, u, 𝖌, λ)$ (note that if $u$ is of the form $\memobraces{u'}[(ƒ, a)][γ]$, then it is assumed that the function--atom label pair $(ƒ, a) ∈ 𝖌_L × 𝖌_R$), we will apply the following reduction rules:

\begin{longtable}{|l c l|}
  \caption{Reduction rules.} \\
  \hline
  \multicolumn{3}{|c|}{\textbf{Reduction Rules}}\\
  \hline
  \endfirsthead
  \hline
  \multicolumn{3}{|c|}{} \\[-1.5ex]
  \endhead
  \multicolumn{3}{|c|}{} \\[-1.5ex]
  \hline
  \multicolumn{3}{|c|}{\textit{Continued on next page}}\\
  \hline
  \endfoot
  \hline
  \endlastfoot
  \hline
  \multicolumn{3}{|c|}{} \\[-1.5ex]
  $(γ, \iletin{x}{\ireturn(v)}{u}, 𝖌, λ)$ & $⟶$ & $(γ ⨆ \{ x ↦ \denpar[γ]{v}\}, u, 𝖌, λ)$ \\
  $(γ, \memobraces{\ireturn(v)}[(ƒ, a)][γ'], 𝖌, λ)$ & $⟶$ & $\begin{cases}
    (γ', \ireturn(\denpar[γ]{v}), (𝖌_L, 𝖌_R, E ∪ \{ ƒ \xto{\denpar[γ]{v}} a \}), λ)\\
    \hspace{7em} \quad\text{ if } \denpar[γ]{v} ∈ \{\itrue, \ifalse\}\\
    \text{ else, failure (cannot memoize a non-boolean function)}\\
  \end{cases}$ \\
  $(γ, \iletin{x}{\lambdamem y u}{t}, 𝖌, λ)$ & $⟶$ & $\big(γ ⨆ \{ x ↦ ƒ\}, t, (𝖌_L ⨆ \{ ƒ\}, 𝖌_R, E ⨆ \{ f \xto {⊥} a \}_{a ∈ 𝖌_R}),$ \\
  && \hphantom{$\big(γ ⨆ \{ x ↦ ƒ\}, t, $} $λ ⨆ \{ ƒ ↦ (\lambdamem y u, γ)\}\big)$ \\
  $(γ, \iletin{x}{\ifresh}{t}, 𝖌, λ)$ & $⟶$ & $\big(γ ⨆ \{ x ↦ a\}, t,$ \\
  && \hphantom{$\big(γ ⨆ \{ x ↦ $} $(𝖌_L, 𝖌_R ⨆ \{ a\}, 𝖌_R, E ⨆ \{ ƒ \xto {⊥} a \}_{ƒ ∈ 𝖌_L}), λ\big)$ \\
  $(γ, (v@w), 𝖌, λ)$ & $⟶$ & $\begin{cases}
    (γ, \ireturn(β), 𝖌, λ) \quad\text{ if } β ≝ E(\denpar[γ]{v}, \denpar[γ]{w}) ≠ ⊥\\
    (γ_0 ⨆ \{ y ↦ \denpar[γ]{w}\}, \memobraces{u}, 𝖌, λ) \quad\text{ else,}\\
    \hspace{7em}\text{where } λ(\denpar[γ]{v}) ≝ (\lambdamem y u, γ_0)
  \end{cases}$ \\
  $(γ, v = w, 𝖌, λ)$ & $⟶$ & $(γ, \ireturn(β), 𝖌, λ) \text{ where } β ≝ (\denpar[γ]{v} = \denpar[γ]{w})$ \\
  $(γ, \iflip, 𝖌, λ)$ &$\overset{\text{\tiny with proba. } θ}{⟶}$& $(γ, \ireturn(β), 𝖌, λ) \quad\text{ where } β ∈ \{ \itrue, \ifalse\}$ \\
  $(γ, \imatch v {(x,y)} t, 𝖌, λ)$ & $⟶$ & $(γ ⨆ \{ x ↦ \denpar[γ]{v}, y ↦ \denpar[γ]{w}\}, t, 𝖌, λ)$ \\
  $(γ, \ite v t u)$ & $⟶$ & $\begin{cases}
    (γ, t, 𝖌, λ) \qquad\text{ if } v = \itrue\\
    (γ, u, 𝖌, λ) \qquad\text{else, if } v = \ifalse
  \end{cases}$ \\
  \multicolumn{3}{|c|}{} \\
  \multicolumn{3}{|c|}{$\dfrac{(γ, e, 𝖌, λ) ⟶ (γ', e', 𝖌', λ')}{(γ, \C[e], 𝖌, λ) ⟶ (γ', \C[e'], 𝖌', λ')}$} \\[1.5em]
  \hline
\end{longtable}

\begin{example} We now give an example showcasing how these reduction rules apply on a program combining name generation, a coin flip, function abstraction, and stochastic memoization. An atom $x_0$ is generated and used as an argument for a function $f_1$, which performs a coin flip if the argument matches $x_0$. The outcome is then memoized and the result is returned in the second application. There are two execution traces, depending on the outcome of the coin flip ($β ∈ {\itrue, \ifalse}$).
  \begingroup
  \allowdisplaybreaks
    \begin{longtable}{p{0.4\textwidth}p{0.1\textwidth}p{0.4\textwidth}}
    $\begin{aligned}
        &\Big(∅, \quad \iletin{x_0}{\ifresh}{\\
        & \hspace{2.6em} \iletin{f_1}{\lambdamem{x}{(\iletin{b}{(x = x_0)}{}\\
        & \hspace{8.5em} \ite b {\iflip[\tfrac 1 2]} {\ifalse})}}{\\
        & \hspace{2.6em} \iletin{f_2}{\lambdamem{y}{f_1 @ y}}{f_2 @ x_0}}},\\
        & \quad (∅, ∅, ∅),\; ∅\Big)
    \end{aligned}$
    & $\hspace{3em} ⟶$ &
    $\begin{aligned}
      &\Big(\{x_0 ↦ a_0\}, \\
      & \hspace{1em}  \iletin{f_1}{\lambdamem{x}{(\iletin{b}{(x = x_0)}{}\\
      & \hspace{6.9em} \ite b {\iflip[\tfrac 1 2]} {\ifalse})}}{}\\
      & \hspace{1em} \iletin{f_2}{\lambdamem{y}{f_1 @ y}}{f_2 @ x_0},\\
      & \quad (∅, \{a_0\}, ∅),\; ∅\Big)
    \end{aligned}$ \\[1em]
    $\begin{aligned}
      ⟶^2  & \Big(\overbrace{\{x_0 ↦ a_0,\, f_1 ↦ ƒ_1,\, f_2 ↦ ƒ_2\}}^{≝ \, γ_0},\quad f_2 @ x_0,\\
      &\hspace{1em} (\{ƒ_1, ƒ_2\}, \;\{a_0\}, \{ƒ_1 \xto {⊥} a_0, \,ƒ_2 \xto {⊥} a_0\}), \\[1em]
      & \quad \big\{ƒ_1 ↦ (\lambdamem{x}{\iletin{b}{(x = x_0)}{}\\
      & \hspace{4.4em} \ite b {\iflip[\tfrac 1 2]} {\ifalse}}), \{x_0 ↦ a_0\}),\\
      & \quad \, ƒ_2 ↦ (\lambdamem{y}{f_1 @ y}, \{x_0 ↦ a_0, f_1 ↦ ƒ_1\})\big\}\Big)
    \end{aligned}$
    & $\hspace{3em} ⟶$ &
    $\begin{aligned}
      & \Big(\overbrace{\{x_0 ↦ a_0,\, f_1 ↦ ƒ_1, \, y ↦ a_0\}}^{≝ \, γ_1},\quad \memobraces{f_1 @ y}[ƒ_2, a_0][γ_0], \\
      & \quad (\{ƒ_1, ƒ_2\}, \;\{a_0\}, \{ƒ_1 \xto {⊥} a_0, \,ƒ_2 \xto {⊥} a_0\}), \\
      & \quad \big\{ƒ_1 ↦ (\lambdamem{x}{\iletin{b}{(x = x_0)}{}\\
      & \hspace{4.4em}\ite b {\iflip[\tfrac 1 2]} {\ifalse}}), \{x_0 ↦ a_0\}), \\
      & \quad \, ƒ_2 ↦ (\lambdamem{y}{f_1 @ y}, \{x_0 ↦ a_0, f_1 ↦ ƒ_1\})\big\}\Big)
    \end{aligned}$ \\[1em]
    $\begin{aligned}
      ⟶  & \Big(\{x_0 ↦ a_0,\, x ↦ a_0\},\\
      & \hspace{.5em} \Bigmemobraces{\bigmemobraces{\iletin{b}{(x = x_0)}{}\\[-.5em]
      & \hspace{4.4em} \ite b {\iflip[\tfrac 1 2]} {\ifalse}}[ƒ_1, a_0][γ_1]}[ƒ_2, a_0][γ_0]\!\!\!, \\
      & \quad (\{ƒ_1, ƒ_2\}, \;\{a_0\}, \{ƒ_1 \xto {⊥} a_0, \,ƒ_2 \xto {⊥} a_0\}), \\
      & \quad \big\{ƒ_1 ↦ (\lambdamem{x}{\iletin{b}{(x = x_0)}{}\\
      &\hspace{4.4em}\ite b {\iflip[\tfrac 1 2]} {\ifalse}}), \{x_0 ↦ a_0\}), \\
      & \quad \, ƒ_2 ↦ (\lambdamem{y}{f_1 @ y}, \{x_0 ↦ a_0, f_1 ↦ ƒ_1\})\big\}\Big)
    \end{aligned}$
    & $\hspace{3em} ⟶$ &
    $\begin{aligned}
      & \Big(\{x_0 ↦ a_0,\, x ↦ a_0, \, b ↦ \itrue\},\\ 
      &\hspace{.5em}\memobraces{\memobraces{\ite b {\iflip[\tfrac 1 2]} {\ifalse}}[ƒ_1, a_0][γ_1]}[ƒ_2, a_0][γ_0]\!\!\!, \\
      & \quad (\{ƒ_1, ƒ_2\}, \;\{a_0\}, \{ƒ_1 \xto {⊥} a_0, \,ƒ_2 \xto {⊥} a_0\}), \\
      & \quad \big\{ƒ_1 ↦ (\lambdamem{x}{\iletin{b}{(x = x_0)}{}\\
      &\hspace{4.4em}\ite b {\iflip[\tfrac 1 2]} {\ifalse}}), \{x_0 ↦ a_0\}), \\
      & \quad \, ƒ_2 ↦ (\lambdamem{y}{f_1 @ y}, \{x_0 ↦ a_0, f_1 ↦ ƒ_1\})\big\}\Big)
    \end{aligned}$ \\[1em]
    $\begin{aligned}
      ⟶  & \Big(\{x_0 ↦ a_0,\, x ↦ a_0, \, b ↦ \itrue\},\\ 
      &\hspace{.5em}\memobraces{\memobraces{\iflip[\tfrac 1 2]}[ƒ_1, a_0][γ_1]}[ƒ_2, a_0][γ_0]\!\!\!, \\
      & \quad (\{ƒ_1, ƒ_2\}, \;\{a_0\}, \{ƒ_1 \xto {⊥} a_0, \,ƒ_2 \xto {⊥} a_0\}), \\
      & \quad \big\{ƒ_1 ↦ (\lambdamem{x}{\iletin{b}{(x = x_0)}{}\\
      &\hspace{4.4em}\ite b {\iflip[\tfrac 1 2]} {\ifalse}}), \{x_0 ↦ a_0\}), \\
      & \quad \, ƒ_2 ↦ (\lambdamem{y}{f_1 @ y}, \{x_0 ↦ a_0, f_1 ↦ ƒ_1\})\big\}\Big)
    \end{aligned}$ 
    & $\hspace{2.3em} \overset{\text{\tiny proba. } \tfrac 1 2}{⟶}$ &
    $\begin{aligned}
      & \Big(\{x_0 ↦ a_0,\, x ↦ a_0, \, b ↦ \itrue\},\\ 
      &\hspace{.5em}\memobraces{\memobraces{\ireturn(β)}[ƒ_1, a_0][γ_1]}[ƒ_2, a_0][γ_0]\!\!\!, \\
      & \quad (\{ƒ_1, ƒ_2\}, \;\{a_0\}, \{ƒ_1 \xto {⊥} a_0, \,ƒ_2 \xto {⊥} a_0\}), \\
      & \quad \big\{ƒ_1 ↦ (\lambdamem{x}{\iletin{b}{(x = x_0)}{}\\
      &\hspace{4.4em}\ite b {\iflip[\tfrac 1 2]} {\ifalse}}), \{x_0 ↦ a_0\}), \\
      & \quad \, ƒ_2 ↦ (\lambdamem{y}{f_1 @ y}, \{x_0 ↦ a_0, f_1 ↦ ƒ_1\})\big\}\Big)
    \end{aligned}$
    \\[1em]
    $\begin{aligned}
      ⟶^2  & \Big(\overbrace{\{x_0 ↦ a_0,\, f_1 ↦ ƒ_1,\, f_2 ↦ ƒ_2\}}^{≝ \, γ_0},\quad \ireturn(β), \quad (\{ƒ_1, ƒ_2\}, \;\{a_0\}, \{ƒ_1 \xto {β} a_0, \,ƒ_2 \xto {β} a_0\}), \\
      & \quad \big\{ƒ_1 ↦ (\lambdamem{x}{\iletin{b}{(x = x_0)}{}\ite b {\iflip[\tfrac 1 2]} {\ifalse}}), \{x_0 ↦ a_0\}), \\
      & \quad \, ƒ_2 ↦ (\lambdamem{y}{f_1 @ y}, \{x_0 ↦ a_0, f_1 ↦ ƒ_1\})\big\}\Big)
    \end{aligned}$
    \end{longtable}
  \endgroup
\end{example}

\paragraph*{Configuration judgements.} We now show that the operational semantics satisfies the memoization equations \cref{eqn:stochmem}. Initial configurations are of the form $(∅, e, ∅, ∅)$, where $e$ is a non-extended expression. One can associate a \emph{configuration judgment} $\J(∅, e, ∅, ∅) ≝ \cje{∅}{∅}{e}{A}$ to every initial configuration. A configuration $(γ, e, 𝖌, λ)$ is said to be \emph{accessible} (from $(∅, e_0, ∅, ∅)$) if there exists a reduction trace $s = (∅, e_0, ∅, ∅) ⟶^\ast (γ, e, 𝖌, λ)$ with probability $> 0$. The big-step operational semantics of an initial configuration is defined in a standard way as the resulting probability distribution over the set of configurations accessible from it. We can then prove that a configuration is accessible only if it has a corresponding configuration judgment that is derivable:

\begin{lemma}
  If a configuration $(γ, e, 𝖌, λ)$ is accessible, there exists a corresponding \emph{configuration judgement} $\J(γ, e, 𝖌, λ) ≝ \cje{Γ}{Δ}{e}{A}$ where $γ ∈ \denpar{Γ}$ and such that $\J(γ, e, 𝖌, λ)$ is derivable (with \cref{fig:language,fig:extended_expressions}).
\end{lemma}

Due to the fact that we have at most one redex per (extended) expression and we do not have recursion (so the dataflow graph does not have self-loops and is acyclic), we can prove that: 

\begin{lemma}
  If a configuration of the form $(γ, \C[v@w], 𝖌, λ)$ is accessible and $E(\denpar[γ]{v}, \denpar[γ]{w}) = ⊥$, then $J(γ, \C[v@w], 𝖌, λ) ≝ \cje{Γ}{Δ}{\C[v@w]}{A}$ is such that the memoization stack $Δ$ does not contain a function--atom label pair with $\denpar[γ]{v}$ as first component.
\end{lemma}

As a corollary, we can then prove that a configuration is accessible only if its memoization stack has no duplicates:

\begin{lemma}
  If a configuration $(γ, e, 𝖌, λ)$ is accessible and $\J(γ, e, 𝖌, λ) ≝ \cje{Γ}{Δ}{e}{A}$ is its corresponding configuration judgment, there is no duplicate in $Δ$.
\end{lemma}

This in turn enables us to ensure that the operational semantics satisfies the memoization equations: 

\begin{proposition}
  If $e_1$ and $e_2$ are programs of the form 
  \begin{align*}
    e_1 &≝ \iletin x {\ifresh} {\iletin{f}{\lambdamem{y}{e}}{\iletin {v_1} {f @ x}{\iletin {v_2} {f @ x}{ \ireturn (v_1, v_2)}}}}\\ 
    e_2 &≝ \iletin x {\ifresh} {\iletin{f}{\lambdamem{y}{e}}{\iletin {v_1} {f @ x}{\ireturn (v_1, v_1)}}}
  \end{align*}
the configurations $(∅, e_1, ∅, ∅)$ and $(∅, e_2, ∅, ∅)$ have the same big-step operational semantics.
\end{proposition}



\section{Denotational Semantics}
\label{sec:denotational-semantics}

In this section we propose a denotational model that verifies the dataflow property (Def.~\ref{eqn:dataflow}, Theorem~\ref{thm:probabilistic-local-state-monad-commutative-affine}) and which supports memoization of constant Bernoulli functions (Theorem~\ref{thm:T-mem}) and is sound with respect to the operational semantics of Section~\ref{sec:operational} (Theorem~\ref{thm:soundness}). Thus we show that criteria (1)--(5) of Section~\ref{sec:intro} are consistent. 

The memo-tables in memoization are a kind of hidden or local state, and our semantic domain is similar to other models of local state \cite{plotkinNotionsComputationDetermine2002,poy-bi,melliesLocalStatesString2014,klms-ground-state-monad} in that it uses a possible worlds semantics in the guise of a functor category.

\begin{definition}
  A \emph{total bigraph} is a partial bigraph (Def.~\ref{def:bigraph}) that does not have any undefined ($\bot$) elements. This represents a fully populated memo-table. We notate this $g=(g_L,g_R,E^g)$, omitting the superscript when it is clear. 
An \emph{embedding} between total bigraphs $\iota\colon g\to g'$ is a pair of injections $(\iota_L: g_L\to g_L',\iota_R: g_R\to g_R'$) that do not add or remove edges ($E^g(ƒ, a)=E^{g'}(\iota_L(ƒ),\iota_R( a))$). 
These can be thought of as conservative extensions of the memo-table.
We let $\BiG$ be the category where the objects are total finite bigraphs and graph embeddings.
\end{definition}

We will interpret our types as covariant presheaves, i.e.~functors in $[\BiG,\Sets]$, and programs will be interpreted as natural transformations. We discuss this category in Section~\ref{sec:presheaf-cat}, before defining a monad (\S\ref{sec:prob-loc-monad}) and giving a denotational semantics (\S\ref{sec:cat-semantics}) and proving a soundness theorem (\S\ref{sec:soundness}). 


\hide{Stochastic functions are usually interpreted as probabilistic kernels $f： X → PY$, where $P$ is a probability monad on a suitable category~\cite{fritz,kock}. A semantic model for stochastic memoization should then admit a cartesian closed structure (to model higher-order functions) and a morphism ${\mathop{\rm mem}}_{X, Y}：(PY)^X ⟶ P(Y^X)$.\todo{This was already said} 
To prove that equations~(\ref{eqn:stochmem}) are consistent with the dataflow property~(\ref{eqn:dataflow}), we give a denotational model. For simplicity, we focus on Boolean-valued functions over a non-enumerable type of atoms~\cite{pittsNominalSetsNames2013}.\todo{This was already said} Our model is based on functors over finite bipartite graphs (\emph{bigraphs}, for short\todo{already said. but emphasise total}), see~\cite{kaddarModelStochasticMemoization2022,kaddarTransferReportExchangeability2022}\todo{Do we need to cite these?} 
  for details. At a lower level, we have the following interface:

\begin{lstlisting}
new_atom :: *'$\mathbb{A}$'* -- Atoms (randomly generated fresh names)
new_function :: *'$\mathbb{F}$'* -- Function labels: type to be thought of as *'$\color{mygray}{\mathbb{A}}$'* -> Bool
(@) :: (*'$\mathbb{F}$'*, *'$\mathbb{A}$'*) -> Bool -- Application operator making every function memoized: type of a bigraph
\end{lstlisting}

where every function from a set of atoms $𝔸$ to \lstinline|Bool| is viewed as an inhabitant of type $𝔽$ (thought of as $𝔸 → \mlstinline{Bool}$). Applying a function to an argument and memoizing the result amounts to the explicit `apply' operator $\mlstinline{(@)}:: 𝔽 × 𝔸 → \mlstinline{Bool}$. But requiring that the results be memoized is precisely saying that $\mlstinline{@}$ ought to be seen as the `edge' relation of a bigraph with set $𝔽$ of left nodes and $𝔸$ of right nodes, the edges of which are such that their presence (or absence) remains unchanged after being sampled.\todo{Some of this has already been said} Inspired by the local state monad~\cite{plotkinNotionsComputationDetermine2002}, we model probabilistic and name generation effects by a new monad $T$ on $[\BiG, \Sets]$, where $\BiG$ is the category of finite bigraphs and embeddings. We then use it to give a categorical semantics to our language, and we show that $T$ is commutative and affine, and hence satisfies the dataflow property~(\ref{eqn:dataflow}).}


\subsection{Base category}\label{sec:presheaf-cat}

We work in the category $[\BiG, \Sets]$ of covariant presheaves on the category $\BiG$ of finite bigraphs. The types of the language $A$ are interpreted as presheaves~$\den A$. The idea is that once some functions and atomic names are fixed, and a memo-table $g$ for them is given, then we can say what the values or expressions are, $\den A(g)$. The values can be renamed by permuting functions and atomic names, and are monotonic in that they remain unchanged when we conservatively extend the memo-table. This is the functorial action, $\den A\iota: \den A(g)\to \den A (g')$. 
Programs will be interpreted as natural transformations: the naturality ensures that they are invariant under permuting the functions and atomic names, or extending the memo-table.

We write $\circ$ and $\bullet$ for the one-vertex left and right graphs respectively. The denotation of basic types is given by:
\[
  \den\Fns ≝ \BiG(\circ,-)\quad
  \den\Atoms ≝ \BiG(\bullet,-)
  \quad\text{so that }
  \den\Fns(g) \cong g_L\text,\
  \den\Atoms(g) \cong g_R\text.
\]
The presheaf category $[\BiG,\Sets]$ has products and coproducts, given pointwise~\cite{cwm}.
In particular, the denotation of the type of booleans is the constant presheaf $2 ≅ 1+1$.

The edge relations collect to form a natural transformation $\mathcal E:\den\Fns\times\den\Atoms\to 2$ given by $\mathcal E_g(ƒ, a) = E^g(ƒ, a)$.

The category $[\BiG,\Sets]$ is cartesian closed, as is any presheaf category. By currying $\mathcal E$,
we have an embedding of $\den\Fns$ in the function space $2^{\den\Atoms}$, i.e.~$\den\Fns\to 2^{\den\Atoms}$. In fact, in this development to keep things simpler, we will focus on $\den\Fns$ rather than the full function space $2^{\den\Atoms}$. 


\subsection{Probabilistic local state monad}
\label{sec:prob-loc-monad}
In the following, $X, Y, Z：\BiG → \Sets$ denote presheaves, $g = (g_L, g_R, E^g), g', h, h' ∈ \BiG$ bigraphs, and $ι, ι'：g ↪ g'$ bigraph embeddings. We will omit subscripts when they are clear from the context.

Let $\Prfin$ be the finite distribution monad:
$\Prfin(X)(g)= \big\{p:X(g)~\to~[0,1]~\big|~ \mathsf{supp}(p)\text{ finite and }\sum_x p(x)=1\big\}$.
By considering the following `node-generation' monad $N(X)(g) ≝ \colim_{g \,↪\, h} X(h)$ on $[\BiG, \Sets]$, one could be tempted to think that modeling name generation and stochastic memoization is a matter of composing these two monads. But this is not quite enough. We also need to remember, in the monadic computations, the probability of a function returning $\itrue$ for a fresh, unseen atom. To do so, inspired from Plotkin and Power's local state monad~\cite{plotkinNotionsComputationDetermine2002} (which was defined on the covariant presheaf category $[\Inj, \Sets]$, where $\Inj$ is the category of ﬁnite sets and injections), we model probabilistic and name generation effects by the following monad, defined using a coend~\cite{cwm}, that we name `probabilistic local state monad':

\begin{definition}[Probabilistic local state monad]\label{def:probabilistic-local-state-monad}
    For all covariant presheaves $X：\BiG → \Sets$ and bigraphs $g ∈ \BiG$:
    \[
    T(X)(g) \,≝\, \bigg(\Prfin \int^{g ↪ h} \Big( X(h)\times [0, 1]^{(h-g)_L}\Big)\bigg)^{[0,1]^{g_L}}
\] 
\end{definition}

The monad $T$ is similar to the read-only local state monad, except that any fresh node can be initialized. Every $λ ∈ [0, 1]^{g_L}$ is thought of as the probability of the corresponding function/left node yielding true on a new fresh atom. We will refer to such a $λ$ as a \emph{state of biases}. The coend `glues together' the extensions of the memo-table that are compatible with the constraints imposed by the current computation. The monad allows manipulating probability distributions over such extensions, while keeping track of the probability of new nodes.

Equivalence classes in $\int^{g ↪ h} X(h)\times [0, 1]^{(h-g)_L}$ are written $[x_h, λ^h]_g$. In the coend, the quotient can be thought of as taking care of garbage collection: nodes that are not used in the bigraph environment can be discarded. We use Dirac's bra-ket notation\footnote{popularized by Bart Jacobs for finite probability distributions~\cite{jacobs-new-directions}} $\Ket{[x_h, λ^h]_g}_h$ to denote a formal column vector of equivalence classes ranging over a finite set of $h$'s. As such, a formal convex sum $\sum_i p_i [x_{h_i}, λ^{h_i}]_g ∈ \Prfin \int^{g ↪ h} X(h)\times [0, 1]^{(h-g)_L}$ will be concisely denoted by $\Braket {\vec p | [x_h, λ^h]_g}_h$.

\begin{definition}[Action of $T(X)$ on morphisms] 
    \[
    T(X)(g \xinjto {ι} g')：\\
    \begin{cases}
    \displaystyle\Big(\Prfin \int^{g ↪ h} \mkern-30mu X(h)\times [0, 1]^{(h-g)_L}\Big)^{[0,1]^{g_L}}
    ⟶ \Big(\Prfin \int^{g' ↪ h'} \mkern-35mu X(h')\times [0, 1]^{(h'-g')_L}\Big)^{[0,1]^{g'_L}} \span\\
    ϑ ↦ [0,1]^{g'_L} \xto {- \circ ι_L} [0,1]^{g_L} &\xto {ϑ} \Prfin \displaystyle\int^{g ↪ h} X(h)\times [0, 1]^{(h-g)_L} \\
    &\xto{\Prfin ψ_{g, g'}} \Prfin \displaystyle\int^{g' ↪ h'} X(h')\times [0, 1]^{(h'-g')_L}
    \end{cases}
    \]
    where 
    
    \begin{itemize}
        \item $ι_L：g_L ↪ g_L'$ is the embedding restricted to left nodes, the maps $ψ_{g, g'}$ are given by:\,\\[1em] 
        \begin{prooftree}
            \hypo{\begin{cases}
                X(h)\times [0, 1]^{(h-g)_L} → X(h \pushout_g g')\times [0, 1]^{(h \pushout_g g' - g')_L} → \mathrlap{\int^{g' ↪ h'} X(h')\times [0, 1]^{(h'-g')_L}}\\
                (x_h, \, λ^h) ⟼ (X(h ↪ h \pushout_g g')(x_h), \, λ^h) ⟼ \mathrlap{[X(h ↪ h \pushout_g g')(x_h), \, λ^h]_{g'}}\vphantom{\bigg|}
            \end{cases}}
            \infer1[extranatural in $h$]{\int^{g ↪ h} X(h)\times [0, 1]^{(h-g)_L} \xto{ψ_{g, g'}} \int^{g' ↪ h'} X(h')\times [0, 1]^{(h'-g')_L}}
        \end{prooftree}\\[1em]
    \item  and $h \pushout_g g'$ is the pushout in the category of graphs regarded as an object of $\BiG$. 
    \end{itemize}
    
    More concretely, with Dirac's bra-ket notation, $T(X)(g \xinjto {ι} g')$ can be written as:
    \[
    T(X)(ι) =
    \begin{cases}
        \Big(\Prfin \int^{g ↪ h} X(h)\times [0, 1]^{(h-g)_L}\Big)^{[0,1]^{g_L}}
        ⟶ \Big(\Prfin \int^{g' ↪ h'} X(h')\times [0, 1]^{(h'-g')_L}\Big)^{[0,1]^{g'_L}} \\
        ϑ ⟼ λ' ↦ \letin{ϑ(λ'ι_L)}{\Braket {\vec p | [x_h, λ^h]_g}_h}{\Braket {\vec p | [X(h ↪ h \pushout_g g')(x_h), λ^h]_{g'}}_h}
    \end{cases}
    \]
\end{definition}

$T$ can be endowed with the structure of a $[\BiG, \Sets]$-enriched monad, that is, since $[\BiG, \Sets]$ is a (cartesian) monoidal closed category, a strong monad. Its enriched unit $η_X：1 → TX^X$ and bind $(-)^\ast：TY^X → TY^{TX}$ are as follows\footnote{following Fosco Loregiàn~\cite{loregian-coend}, $よ：\BiG → [\BiG, \Sets]\op$ denotes the (contravariant) Yoneda embedding.}.

\begingroup
\allowdisplaybreaks
\begin{longtable}{p{0.63\textwidth}p{0.37\textwidth}}
$\begin{aligned}
  η_X(g)：\begin{cases}
      \{*\} ⟶ [X × よ(g), TX]\\
      * ⟼  ŋg'：
      \begin{cases}
      X(g') × よ(g)(g') ⟶ TX(g') \\
      x_{g'}, \; \text{\textunderscore} ⟼ \Big([0, 1]^{g'_L} ∋ λ ↦ 1 ⋅ \Ket{[x_{g'}, !]_{g'}}\Big)
  \end{cases} 
  \end{cases}
  \end{aligned}$ &\hspace{-2
  em}$\begin{aligned}
  (-)^\ast：\begin{cases}
  TY^X(g) ⟶ \big[TX × よ(g), TY\big]
  \\
  φ ⟼ φ^\ast
  \end{cases}
  \end{aligned}$\\
\text{where } \\[-.2em]
\multicolumn{2}{l}{
$\begin{aligned}
  φ^\ast_{g'}： \begin{cases}
    \displaystyle\Big(\Prfin \int^{g' ↪ h} \mkern-35mu X(h)\times [0, 1]^{(h-g')_L}\Big)^{[0,1]^{g_L'}} \mkern-15mu × よ(g)(g') \xto {φ^\ast_{g'}} \Big(\Prfin \int^{g' ↪ h'} \mkern-35mu Y(h')\times [0, 1]^{(h'-g')_L}\Big)^{[0,1]^{g_L'}}
    \span\\
    (ϑ, \; g \xinjto {ι} g') \xmapsto {φ^\ast_{g'}} λ' ↦& \letin {ϑ(λ')} {\Braket {\vec p | [x_h, λ^h]_{g'}}_{h ∈ H_{g'}}}{\\
    &\for{h ∈ H_{g'}}\\
    & \;\; \letin{φ_h(x_h, g \xinjto {ι} g' ↪ h)(λ^h ⨆ λ')}{\Braket {\vec {q_h} | [y_h', γ^{h'}]_{h}}_{h' ∈ H_{h}}}{\\
    & \Braket {\vec p | Q | [y_{h'}, γ^{h'} ⨆ λ^h]_{g'}}_{h' ∈ \bigcup\limits_{h ∈ H_{g'}} \mkern-15mu H_h}}}\\
\end{cases}
\end{aligned}$}\\
\text{and }
$Q \,≝\, \begin{pmatrix}
    \vec {q_{h_1}} \\
        \vdots \\
        \vec {q_{h_n}}
\end{pmatrix}_{{\smash h_i ∈ H_{g'}}}
\hspace{-1em}=\hspace{1em}
\begin{pmatrix}
    q_{h_1, h_1'}   & \rvline &   & \rvline &  q_{h_1, h_m'} \\
    \vdots          & \rvline & ⋯ & \rvline & \vdots \\
    q_{h_n, h_1'}   & \rvline &   & \rvline &  q_{h_n, h_m'}
    \end{pmatrix}_{\substack{h_i ∈ H_{g'} \\ h_j' ∈ \bigcup\limits_{h ∈ H_{g'}} \mkern-15mu H_h}}$ & \hspace{-4.5em}\text{(where each $\vec {q_h}$ has been $0$-padded accordingly)}\\
\end{longtable}
\endgroup

As argued before, to construct an abstract model of probability, we show that the monad is commutative. Affineness straightforwardly stems from the following lemma:

\begin{lemma}\label{lemma:T-on-constant-presheaves}
  Let $X$ be a constant presheaf on the coslice category $g/\BiG$, \ie there exists a set $S_0$ such that
  $X(g \xinjto {ι} h) = S_0 \xto {\id} S_0$
  for every $g \xinjto {ι} h \in g/\BiG$. Then $T(X)(g) ≅ \Prfin(S_0)^{[0, 1]^{g_L}}$.
\end{lemma}

We have the desired dataflow property, meaning that $T$ is an abstract model of probability~\cite{kock}:
\begin{theorem}
  \label{thm:probabilistic-local-state-monad-commutative-affine}
  The monad $T$ satisfies the dataflow property~(\ref{eqn:dataflow}): it is strong commutative and affine. 
\end{theorem}

\begin{proof*}{Proof (Sketch)}
In the presheaf category, let $Z^Y × Y^X \xto {\circ} Z^X$ and $Z^Y × Y \xto {\rm ev} Z$ denote the internal composition and evaluation, and $f^\ast ≝ 1 \xto f TY^X \xto {(-)^\ast} TY^{TX}$ the internal Kleisli lifting of a global element $f$. To prove that $T$ is strong, we show, internally, the associativity ($(Ψ^\ast_g × Φ^\ast_g) \then \circ = ((Ψ^\ast × Φ) \then \circ)^\ast$) of the bind, the left unit law ($η^\ast = λ_{TX}. \id_{TX}$), and the right unit law ($(Φ^\ast × η) \then \circ = Φ$), for all $Φ：1 → TY^X, Ψ：1 → TZ^Y$. Finally, affineness stems from lemma~\ref{lemma:T-on-constant-presheaves}, and commutativity is the equation $a \bind \lambdaabs{x}{b \bind \lambdaabs{y}{η(x, y)}} \;=\; b \bind \lambdaabs{y}{a \bind \lambdaabs{x}{η(x, y)}}$ internally, for all $a：1 → TA, b：1 → TB$, which amounts to showing: $$\bigg(\Big(λ_A. \Big(\big((λ_B. η)^\ast × b\big) \,\then\, {\rm ev}\Big)\Big)^\ast × a\bigg) \then {\rm ev} = \bigg(\Big(λ_B. \Big(\big((λ_A. η)^\ast × a\big) \,\then\, {\rm ev}\Big)\Big)^\ast × b\bigg) \then {\rm ev}$$
\end{proof*}

\subsection{Categorical semantics}
\label{sec:cat-semantics}

In our language, the denotational interpretation of values, computations (return and let binding), and matching (elimination of $\tbool$'s and product types) is standard. We interpret computation judgements $\cj {Γ} t A$ as morphisms
$\den {Γ} \to T (\den A)$, by induction on the structure of typing derivations. The context $Γ$ is built of $\tbool$'s, $\Fns$, $\Atoms$ and products. Therefore, $\den {Γ}$ is isomorphic to a presheaf of the form $2^k × \BiG(\circ, -)^\ell × \BiG(\bullet, -)^m$, where $k, \ell, m$ are the numbers of booleans, functions and atoms in $Γ$, and $X^n$ is is the $n$-fold finite product in the category of presheaves. Computations of type $\Atoms$ and $\Fns$ then have an intuitive interpretation:

\begin{proposition}\label{propos:denotation-atoms-functions}
  A computation of type $\Atoms$ returns the label of an already existing atom or a fresh one with its connections to the already existing functions: $T(\den \Atoms)(g) \,≅\, \Prfin(g_R + 2^{g_L})^{[0, 1]^{g_L}}$. A computation of type $\Fns$ returns the label of an already existing function or create a new function with its connections to already existing atoms and a fixed probabilistic bias: $T(\den \Fns)(g) \,≅\, \Prfin(g_L + 2^{g_R} × [0, 1])^{[0, 1]^{g_L}}$.
\end{proposition}

For every bigraph $g$, we denote by $R_g$ (resp. $L_g$) the set of bigraphs $h ∈ g/\BiG$ having one more right (resp. left) node than $g$, and that are the same otherwise. For every $e ∈ 2^{g_L}$ (resp. $e ∈ 2^{g_R}$), we denote by $g +_e \bullet ∈ R_g$ (resp. $g +_e \circ ∈ L_g$) the bigraph obtained by adding a new right (resp. left) node to $g$ with connectivity $e$ to the right (resp. left) nodes in $g$. We now give the denotational semantics of various constructs in our language. Henceforth, we will denote normalization constants (that can easily be inferred from the context) by $Z$.

\paragraph*{Denotations of $\cj{Γ}{\iflip} \tbool$, $\cj{Γ, v: \Fns, w: \Atoms}{v @ w} \tbool$, and $\cj{Γ, v: \Atoms, w: \Atoms}{v = w} \tbool$} First, by Lemma~\ref{lemma:T-on-constant-presheaves}, we note that $T(\den \tbool)g\, ≅ \, \Prfin(2)^{[0, 1]^{g_L}} \, ≅ \, [0, 1]^{[0, 1]^{g_L}}$. So naturally, the map $\den\iflip_g$ is the constant function returning the bias $θ$. 

\paragraph*{Denotations of $\cj{Γ, v: \Fns, w: \Atoms}{v @ w} \tbool$, and $\cj{Γ, v: \Atoms, w: \Atoms}{v = w} \tbool$}

The map $\den{v @ w}_g:\den {Γ, v: \Fns, w: \Atoms}(g) \to [0, 1]^{[0, 1]^{g_L}}$ returns $1$ if the left node corresponding to $v$ is connected to the one of $w$ in $g$, $0$ otherwise. Using the internal edge relation $\mathcal E$, it is the internal composition: $$\den{v @ w} ≝ 1 × (\den{Γ} × \den{𝔽} × \den{𝔸}) \xto {η × (! × \mathcal E)} T(\den{\tbool})^{\den{\tbool}} × \den{\tbool} \xto {{\rm ev}} T(\den\tbool)$$

And similarly, the map $\den{v = w}_g:\den {Γ, v: \Atoms, w: \Atoms}(g) \to [0, 1]^{[0, 1]^{g_L}}$ is given by: $$\den{v = w} ≝ \den{Γ} × \den{𝔸}^2 ≅  1 × \den{Γ} × \Big(よ(\bullet) + よ(\bullet + \bullet)\Big) \xto {η × ! × \big[! \,\then\, ι_{\itrue},\; ! \,\then\, ι_{\ifalse} \big]} T(\den{\tbool})^{\den{\tbool}} × \den{\tbool} \xto {{\rm ev}} T(\den\tbool)$$

where $[-,\, -]$ is the copairing and $ι_{\itrue}, ι_{\ifalse}： 1 → \den\tbool ≅ 2$ are the coprojections.


\paragraph*{Denotation of $\cj{Γ}{\ifresh} \Atoms$.}

The map $\den\ifresh_g:\den {Γ}(g) \to T (\den \Atoms)(g)$ randomly chooses connections to each left node according to the state of biases, and makes a fresh right node with those connections. 

\begin{equation*}\label{eq:denotation-fresh}
\begin{split}
  &\den\ifresh_g：\begin{cases}
    2^k × \BiG(\circ, g)^\ell × \BiG(\bullet, g)^m ⟶ \Prfin(g_R + 2^{g_L})^{[0, 1]^{g_L}} \\
    \text{\textunderscore},\, \text{\textunderscore},\, \text{\textunderscore} ↦ λ ↦ \displaystyle\Braket{\frac 1 Z \smash{\prod\limits_{ƒ ∈ g_L} λ(ƒ)^{E^h(ƒ, a_h(\bullet))} (1-λ(ƒ))^{1-E^h(ƒ, a_h(\bullet))}} | \big[ \smash{\underbrace{\bullet}_{\mathclap{\; ≅ \, (h-g)_R}}} \xinjto {a_h} h, \, !\big]_g}_{h ∈ R_g}
  \end{cases}
\end{split}
\end{equation*}

It suffices to consider the bigraphs that belong to $R_g$ only, by garbage collection of the coend.

\paragraph*{Denotation of $\cj{Γ}{\lambdamem x u} \Fns$.}

As $\lambdamem{}{}$-abstractions are formed based on computation judgements of the form $\cj {Γ,x:\Atoms} u \tbool$. We can decompose the extra variable $x$ in the environment $Γ, \, x:\Atoms$, the denotation of which is of the form $\den{Γ, \, x:\Atoms}(g) \, = \, 2^k × \BiG(\circ, g)^\ell × \BiG(\bullet, g)^m \, × \, \BiG(\bullet, \, g)$ for a bigraph $g ∈ \BiG$. Now, the extra part $x$ is a right node, and its valuation will either be a node already in the graph described in the rest of the environment, or a new one with particular edges to the rest of the environment. The argument $u$ can test (if it wants) what kind of node $x$ is, before returning a probability. 

As a result, the denotation $\den u_g：2^k × \BiG(\circ, g)^\ell × \BiG(\bullet, g)^m × \BiG(\bullet, g) ⟶ [0, 1]^{[0, 1]^{g_L}}$ gives us the edge probability of the left node (atom) that we need to generate, both to the existing right nodes (functions), and to any future right node (which needs to be remembered). This can be formalized into a natural transformation $\den{\lambdamem x u}：\den {Γ}\to T(\den \Fns)$, provided that $u$ satisfies the following property:

\begin{definition}[Freshness-invariant functions]\label{def:freshness-invariant}
  A function $\lambdamem{x}{u}$ is \emph{freshness-invariant} if, for every $g, b^k ∈ 2^k$, $κ_i：\circ ↪ g, \, τ_j ：\bullet ↪ g$ and $λ ∈ [0, 1]^{g_L}$, we have (where $ι_1, \, ι_2$ are the coprojections):
  $$∀ e ∈ 2^{g_L}, \, \den{u}_g\big(b^k,\, (\circ \xinjto {κ_i} g \xinjto {ι_1} g +_e \bullet)_i,\, (\bullet \xinjto {τ_j} g \xinjto {ι_1} g +_e \bullet)_j, \, \bullet \xinjto {ι_2} g +_e \bullet, \, λ\big) \text{ is a constant } \tilde{p}_u$$
\end{definition}

A sufficient condition to ensure that a function of the form $\lambdamem{x}{u}$ be freshness-invariant is that it has no subexpression of the form $f @ y$, where $y ∉ \fv(\lambdamem{x}{u})$. An example thereof is $\lambdamem{x}{\iletin{b}{f @ x_0}{\ite{b}{\itrue}{(x=x_0)}}}$. Non examples are $\lambdamem{x}{\iletin{y}{\ifresh}{f @ y}}$ and $\lambdamem{x}{\ite{f @ x}{\ifalse}{\itrue}}$ (negation of $f$). 
We can interpret freshness-invariant functions as follows: 

\begin{equation*}
  \mkern-20mu\den{\lambdamem x u}_g：\!\!\begin{cases}
      2^k × \BiG(\circ, g)^\ell × \BiG(\bullet, g)^m ⟶ \Prfin(g_L + 2^{g_R} × [0, 1])^{[0, 1]^{g_L}}\\[1em]
      b^k,\, \substack{\mbox{\normalsize $(\circ \xinjto {κ_i} g)_i,$}\\ \mbox{\normalsize $(\bullet \xinjto {τ_j} g)_j$}} ↦ λ ↦ \displaystyle\Braket{\frac 1 Z \smash{\prod\limits_{a ∈ g_R} p_a^{E^h(ƒ_h(\circ), a)} (1-p_a)^{1-E^h(ƒ_h(\circ), a)}} | \big[ \smash{\underbrace{\circ}_{\mathclap{\, ≅ \, (h-g)_L}}} \xinjto {ƒ_h} h, \, \text{\textunderscore} ↦ \tilde{p}_u\big]_g}_{h ∈ L_g}
    \end{cases}\\
    \label{eq:denotation-lambdamem}
\end{equation*}
where $p_a \,≝\, \den{u}_g\big(b^k,\, (\circ \xinjto {κ_i} g)_i,\, (\bullet \xinjto {τ_j} g)_j, \, \bullet \xinjto a g, \, λ\big)$ for every $a ∈ g_R$, and $\tilde{p}_u$ is as in Def.~\ref{def:freshness-invariant}. As a result, the probabilistic local state monad validates~\eqref{eqn:stochmem}:

\begin{theorem}\label{thm:T-mem}
  The monad $T$ supports stochastic memoization (Def.~\ref{def:mem}) for freshness-invariant functions (Def.~\ref{def:freshness-invariant}), which include any function $\lambdamem{x}{u}$ that does not contain a subexpression of the form $f @ y$, where $y ∉ \fv(\lambdamem{x}{u})$ (so, in particular, constant Bernoulli functions).
\end{theorem}
\begin{proof*}{Proof (Sketch)} The denotation of $\lambdamem{}{}$-abstractions enables us to define a map $T(\den\tbool)^{\den{\Atoms}} → T(\Fns)$, which can in turn be postcomposed by $T(\Fns) \xto {φ} T\big(\den{\tbool}^{\den{\Atoms}}\big)$, where

$$
φ_g ：\begin{cases}
  T(\Fns)(g) ≅ \Prfin(g_R + 2^{g_L})^{[0, 1]^{g_L}} ⟶ [0, 1]^{[0, 1]^g × (g_R + 2^{g_L})} ≅ T\big(\den{\tbool}^{\den{\Atoms}}\big)(g)\\
  ϑ ⟼ (λ, a) ∈ [0, 1]^g × (g_R + 2^{g_L})
  ↦ \letin{ϑ(λ)}{\sum_{a' ∈ g_R + 2^{g_L}} p_{a'} \Ket{a'}}{p_a}
\end{cases}
$$

to obtain $\mlstinline{mem}：T(\den\tbool)^{\den{\Atoms}} → T(\den{\tbool}^{\den{\Atoms}})$, and then we show \cref{eqn:mem-monad} in the presheaf topos.
\end{proof*}

\begin{example}
  The denotation of $\iletin{x}{\ifresh}{
    \iletin{f}{\lambdamem{y}{\iflip}}{f @ x}}$ 
  is the map $$1 × 1 \xto {\Big(λ_{\den \Atoms}. \Big(\big((λ_{\den \Fns}. f @ x)^\ast × (\lambdamem{y}{\iflip}) \big) \then {\rm ev}\Big) \Big)^\ast × \, \ifresh} T(\den\tbool)^{T\den\Atoms} × T(\den\Atoms) \xto {{\rm ev}} T(\den\tbool)$$
  given by $\ast, \ast ↦ λ ↦ θ \Ket{\itrue} + (1-θ) \Ket{\ifalse}$, as desired. 
\end{example}

\subsection{Soundness}
\label{sec:soundness}

Configurations are of the form $(γ, e, 𝖌, λ)$, where $e$ is of type $A$, and can be denotationally interpreted as

$$\den{(γ, e, 𝖌, λ)} ≝ \sum_{{\tilde e} ∈ 2^{U_{𝖌}}} \prod_{(ƒ, a) ∈ U_{𝖌}} λ(ƒ)^{{\tilde e}(ƒ, a)} \big(1-λ(ƒ)\big)^{1-{\tilde e}(ƒ, a)} \den[𝖌_{\tilde e}]{u}(γ, λ) ∈ T(A)_{𝖌_{\tilde e}}(γ)(λ)$$

where $U_{𝖌} ≝ \big\{(ƒ, a) \mid E(ƒ, a) = ⊥ \big\} ⊆ 𝖌_L × 𝖌_R$ and $𝖌_{\tilde e}$ extends $𝖌$ according to $\tilde e$: $E(ƒ, a) = {\tilde e}(ƒ, a)$ for all $(ƒ, a) ∈ U_{𝖌}$. We can then prove that the denotational semantics is sound with respect to the operational semantics:

\begin{theorem}[Soundness]\label{thm:soundness}
  $$\den{(γ, e, 𝖌, λ)} ≅  
  \sum_{\substack{(γ, e, 𝖌, λ) ⟶ (γ', e', 𝖌', λ')\\ \text{with proba. } p}} p \cdot \den{(γ', e', 𝖌', λ')}$$
\end{theorem}
\begin{proof*}{Proof (Sketch)} As an intermediate step, we build a big-step semantics, and show that this is sound, \ie~making a small step of the operational semantics (\S\ref{sec:operational}) does not change the distributions in the final big-step semantics. Next, we show that the big step semantics of a configuration corresponds to the denotational semantics, for which the main thing to check is that the equivalence classes of the coend are respected. 
\end{proof*}

\section{Haskell Implementation}
\label{sec:implementation}
We have a practical Haskell implementation comparing the small-step, big-step operational, and denotational semantics to showcase the soundness theorem with QuickCheck, in a setting analogous (albeit slightly different\footnote{Unlike our mathematical framework, where we can memoize all freshness-invariant functions (\ref{def:freshness-invariant}), our implementation only memoizes constant Bernoulli functions. Another key difference is that we could not implement coends in Haskell, so we used a global state monad transformer to manage the memoization bigraph, keeping track of edges between left nodes (function labels) and right nodes (atom labels) that have been sampled.}, to better suit the specificities of Haskell) to the theoretical one we presented. The artefact is openly available~\cite{stoch-mem-implementation-github}.

\section{Summary}

In conclusion, we have successfully tackled the open problem of finding a semantic interpretation of stochastic memoization for a class of functions with diffuse domain that includes the constant Bernoulli functions. Our contributions pave the way for further exploration and development of probabilistic programming and the sound application of stochastic memoization in Bayesian nonparametrics.

\section{Acknowledgements}
We are grateful to Nate Ackerman, Cameron Freer, Dan Roy and Hongseok
Yang for various conversations over many years, relating
to~\cite{statonExchangeableRandomProcesses2017}, name generation, stochastic memoization and subsequent developments.
The presheaf category here is related to the Rado
topos~\cite{caramello-fraisse} that we have been exploring in ongoing
work, with Jacek Karwowski and Sean Moss and the above four coauthors.
Thanks to Dario Stein for discussions about name generation and for
pointing out~\cite{kallianpur}.
Thanks too to Swaraj Dash, Mathieu Huot, Ohad Kammar, Oleg Kiselyov, Alex Lew, and all in the Oxford group for many discussions about this topic. 
Finally, thank you to our reviewers for detailed feedback. 

\bibliographystyle{./entics}
\bibliography{./bibliography}

\end{document}